\documentclass[aps,pra,preprint]{revtex4-2}
\usepackage{amsmath, amssymb, graphicx,hyperref}

\newcommand{\nn}{\nonumber}
\newcommand{\Gy}{\left(\omega_{1}^2 \tau_c + \omega^2_{\rm SL} \tau_c\right)}

\newcommand{\Gz}{\left(\omega_{1}^2 \tau_c + 2 \, \omega^2_{\rm SL} \tau_c\right)}

\newcommand{\Gbar}{\left(\omega_{1}^2 \tau_c + \tfrac{3}{2} \, \omega^2_{\rm SL} \tau_c\right)}

\begin{document}
\title{Quantum Speed Limit and Optimal Evolution Time in Driven-dissipative Systems}

\author{Sarfraj Fency$^{1}$\footnote{smjf21ip029@iiserkol.ac.in}, 
Riddhi Chatterjee$^{2}$\footnote{criddhi@iisc.ac.in}, and Rangeet
Bhattacharyya$^{1}$\footnote{rangeet@iiserkol.ac.in}}

\affiliation{$^{1}$Department of Physical Sciences, 
Indian Institute of Science Education and Research Kolkata, Mohanpur 741246, India\\
$^{2}$Department of Instrumentation \& Applied Physics, 
Indian Institute of Science, Bengaluru 560012, India}

\begin{abstract}

Quantum Speed Limit (QSL) is a lower bound on the evolution time $T_{\rm QSL}$ of a closed
quantum system to evolve from one state to another, as shown by Mandelstam and Tamm. For a
driving Hamiltonian enforcing the evolution, the time-energy uncertainty relations for the
drive play the dominant role in setting the bound. On the other hand, dissipative systems
can also suffer from drive-induced dissipation (DID). In this work, we ask how DID
modifies the attainability of the QSL in open quantum systems and whether the time-energy
uncertainty relation remains the dominating factor. For open quantum systems that suffer
from DID, we report a finite optimal evolution time $T_{\rm DID} \; (> T_{\rm QSL})$ for
constant-amplitude driving. We show that the competition between DID at short times and
environmental dissipation at long times produces a non-monotonic dependence of fidelity on
the evolution time. When the drive waveform is shaped using quantum optimal control (QOC),
an intermediate attainable evolution time $T_{\rm QOC}$ provides the maximum fidelity. Our
results therefore suggest that, in driven-dissipative systems, the conventional QSL lower
bound should be complemented by a finite-time attainability window determined by
dissipation, with $T_{\rm QSL} \le T_{\rm QOC} \le T_{\rm DID}$ for the cases studied
here. Further, we show that the optimized control protocol is robust against detuning and
variations in environmental parameters, enabling reliable quantum control in open systems
and providing a practical route toward high-fidelity state transfer in two-level quantum
systems.

\end{abstract}

\maketitle

\section{Introduction} \label{Sec: Introduction}

The quantum speed limit (QSL) time sets a fundamental bound on how fast a quantum system
can evolve between two distinguishable states. The bound is
closely related to the time-energy uncertainty relation \cite{deffner2017quantum}.
The first analytical expression of the bound was calculated by
Mandelstam and Tamm in terms of energy variance $\left(T_{\rm QSL}^{\text{MT}} = \frac{\pi \hbar}{2 \, \Delta E} \right)$
\cite{mandelstam1945uncertainty}, which was later refined by Margolus and Levitin in terms
of mean energy $\left(T_{\rm QSL}^{\text{ML}} = \frac{\pi \hbar}{2 \, \langle E
	\rangle}\right)$ \cite{margolus1998maximum}. The maximum of the two sets a fundamental bound for a closed quantum system \cite{giovannetti2003quantum}. The QSL time establishes
a lower bound on the time required by a quantum system to reach a target state under a
specific dynamical resource. The QSL bounds were originally derived for closed quantum
systems, but they have been generalized to open quantum systems using the quantum master equation \cite{deffner2013quantum}, quantum Fisher information \cite{taddei2013quantum}
and relative purity \cite{del2013quantum}. These bounds play an important role when
operational speed competes directly with decoherence and noise, such as in quantum
information processing \cite{epstein2017quantum, deffner2020quantum, mohan2022quantum},
quantum thermodynamics \cite{campaioli2017enhancing, funo2019speed}, and quantum metrology
\cite{maleki2023speed, herb2024quantum}.

Moreover, in practical scenarios, the interaction with the environment is inevitable, which causes decoherence and introduces noise to the system \cite{breuer2002theory}. But in order to perform quantum operations faster than the decoherence timescale, we need to have
shorter drive period, which requires applying a strong drive with high power. As recently
shown by Chanda \textit{et al.}, stronger drive introduces drive-induced dissipation (DID)
which degrades the population transfer efficiency and limits the performance of the quantum
gate operation \cite{chanda2020optimal, chanda2023optimal}. The use of a recently developed fluctuation regularized quantum master equation (FRQME) allows us to calculate the closed form of DID \cite{frqme}. The standard quantum master equations
consider dissipation originating only from the environment, while the other interactions,
like the external drive, are considered only in the first order \cite{bloch53, redfield,
	lindblad}. But FRQME captures DID by using the bath correlation time ($\tau_c$) to
regularize second-order contributions from both the drive and the system environment
coupling. These features of FRQME are appropriately used in quantum information
processing \cite{chanda2020optimal, chanda2023optimal, saha2024applications}, quantum
optics \cite{chatterjee2024improved}, quantum measurement \cite{chanda2021emergence,
das2024irreversibility} and non-equilibrium dynamics \cite{saha2023cascaded,
saha2024prethermalization, saha2024prethermal, das2025discrete} to describe the dynamics
of driven dissipative quantum systems.

In this work, we ask how the DID modifies the attainability of the QSL in open quantum
systems and whether the time-energy uncertainty relation remains the dominating factor. We
note that the QSL is a \emph{lower} bound on the time needed to reach a given target
state. Here, we show that DID introduces an additional restriction on the attainable
evolution time. Because the drive amplitude required to complete a transfer in a time
$T_{\rm pulse}$ grows as $T_{\rm pulse}$ shrinks, and because DID grows with the drive
amplitude, the fidelity reduces at short $T_{\rm pulse}$; conversely, environmental
relaxation degrades the fidelity at long $T_{\rm pulse}$.  Consequently, for any target
fidelity $\epsilon$ the set of admissible durations is a closed interval
$[T_{\min}(\epsilon),\, T_{\rm max}(\epsilon)]$, which is empty below a
dissipation-imposed floor $\epsilon^{\star}$. The lower edge of this window is the
QSL-type bound while its upper edge and intermediate optimum point arise from the
competition between DID and environmental dissipation. In this precise sense, the
conventional QSL lower bound is complemented by a finite-time feasibility window in a
driven-dissipative system.  We first establish the existence of $T_{\rm DID}$ analytically
in a simple toy model, where it can be obtained in closed form.  A rigorous validation is
then performed using quantum optimal control theory, and we finally decompose the
analytical bound to identify which of its terms dominates at $T_{\rm QSL}$.

The quantum optimal control (QOC) theory provides a systematic framework for engineering
external fields that drive quantum systems toward desired objectives with maximum accuracy
\cite{nielsen2006optimal, werschnik2007quantum}. Among numerical techniques, the Gradient Ascent Pulse Engineering (GRAPE) algorithm has become a standard method for constructing
high-performance pulse sequences under realistic constraints \cite{khaneja2005optimal,
schulte2011optimal}. The gradient-based optimization approach refines the pulse profile by
iteratively maximizing accuracy and minimizing operation time. 

The conceptual bridge between QSLs and QOC was made explicit in seminal work by Caneva
\textit{et. al.}, who demonstrated that optimal control techniques can be used to
numerically determine minimal evolution times that match with analytical QSL bounds for
standard state-transfer problems \cite{caneva2009optimal}. Their results established that,
in unitary settings, optimal control protocols can match fundamental speed limits, thereby
providing an operational interpretation of QSLs as the true boundary of controllability.
Subsequent studies reinforced this connection by showing that time-optimal gate synthesis
and state preparation tasks often approach geometric or energy-based QSL bounds when
sufficient control resources are available \cite{poggi2013quantum, brouzos2015quantum,
PhysRevA.97.062346, cao2024robust, impens2025approaching}. Importantly, however, these
investigations have largely focused on closed quantum systems governed by unitary
dynamics. The connection between QOC and QSL in open quantum systems has been significantly less explored. It is a priori unclear whether numerically optimized protocols can approach, match, or appear to exceed analytical QSL bounds once full
open-system dynamics are consistently incorporated.

The manuscript is organized into different sections. In section \ref{Sec: FRQME}, we give a
brief description of FRQME, which is the mathematical formalism on which our analysis is
based. The details about the model and dynamical equation are given in section \ref{Sec:
Model}. The section \ref{Sec: GRAPE} shows how to implement the GRAPE algorithm in a driven
dissipative quantum system, followed by the numerical results presented in section
\ref{Sec: results}. Finally, the discussion is presented in section \ref{Sec: discussion}
followed by a conclusion.


\section{fluctuation regularized quantum master equation} \label{Sec: FRQME}
In 2018, Chakrabarti \textit{et al.} gave a new formalism to analyze the open quantum systems called fluctuation regularized quantum master equation (FRQME) \cite{frqme}. Here, we give a brief description of FRQME, which is the mathematical formalism on which our analysis is based.

The total Hamiltonian in the laboratory frame is given as follows: 

\begin{equation}
\mathcal{H}(t) = \mathcal{H}^{\circ}_{\text{S}} + \mathcal{H}^{\circ}_{\text{L}} + \mathcal{H}_{\rm SL} + \mathcal{H}_{\text{S}}(t) + \mathcal{H}_{\text{L}}(t),
\end{equation}

where $\mathcal{H}^{\circ}_{\text{S}}$ and $\mathcal{H}^{\circ}_{\text{L}}$ represent the static Hamiltonians of the quantum system and its local-environment, respectively, which are coupled by the term $\mathcal{H}_{\rm SL}$. $\mathcal{H}_{\text{S}}(t)$ denotes the external drive applied on the quantum system while $\mathcal{H}_{\text{L}}(t)$ is the fluctuation Hamiltonian of the local-environment. We transform the total Hamiltonian in interaction picture of $\mathcal{H}^{\circ}_{\text{S}} + \mathcal{H}^{\circ}_{\text{L}}$ which takes the following form:
\begin{equation}
H = H_{\text{S}} + H_{\text{L}} + H_{\rm SL} 
\end{equation}

The thermal noise from the environment was chosen to be diagonal in the eigenbasis of the static Hamiltonian of the environment:
\begin{eqnarray}
\mathcal{H}_{\text{L}} = \Sigma_j f_j \vert \phi_j \rangle \langle \phi_j \vert, 
\end{eqnarray}
here $f_j$ is assumed to be Gaussian, $\delta$-correlated stochastic variable with zero mean and standard deviation $\kappa$. This ensures that the equilibrium population distribution of the environment does not change. 

A finite propagator, which is infinitesimal in terms of the system Hamiltonian but finite in terms of the thermal fluctuations, has been constructed from the Schr{\"o}dinger equation to arrive at the regulator from the thermal fluctuations. We assume that the system and the fluctuations have widely separated timescales, i.e., $\tau_c \ll \tau_s$, where $\tau_c$ is the bath correlation time and $\tau_s$ is the system evolution time.

The mathematical form of the master equation in the interaction picture of the free Hamiltonian is given as: 
\begin{eqnarray} \label{frqme_lab}
\dot{\rho}_s &=& - \; i \: \text{Tr}_{\text{L}} [H_{\text{eff}}, \: \rho]^{\text{sec}} \nn \\ 
&-& \int_{0}^{\infty} d\tau \: e^{-\frac{\tau}{\tau_{c}}} \: 
\text{Tr}_{\text{L}} [H_{\text{eff}}(t),\: [H_{\text{eff}}(t - \tau),\: \rho]]^{\text{sec}}
\end{eqnarray}

Here, $ \tau_{c} = \frac{2}{ \kappa^2}$, $\rho $ is the total density matrix, $\rho_{\rm s} =  \text{Tr}_{\text{L}} (\rho)$ is the density matrix of the system. The superscript ``sec" represents a secular approximation, where only the slow oscillating terms are retained \cite{secular_approx}. To study the system dynamics, a partial trace is taken over the bath degrees of freedom, which is represented as  $\text{Tr}_\text{L}$.

The effective Hamiltonian consists of the drive Hamiltonian as well as the system-bath coupling Hamiltonian, i.e., $H_{\text{eff}} = H_{\text{S}} + H_{\rm SL}$. We note that the sole effect of $H_{\rm SL}$ is only in the dissipator \textit{i.e.} $\text{Tr}_{\text{L}}[H_{\rm SL}, \rho] = 0$.


Further, Born approximation is also used which tells that at the beginning of coarse grain interval, the total density matrix can be factorised into the system and the environment part, i.e. $\rho = \rho_{\rm s} \otimes \rho_{\text{L}}^{eq}$ with $\rho_{\text{L}}^{eq}$ being the equilibrium density matrix of the environment \cite{breuer2002theory}. The second term of FRQME predicts the presence of DID, which comes from the external drive Hamiltonian, which has been verified experimentally \cite{didexpt}.

\section{The Model} \label{Sec: Model}

We consider a spin-$\frac{1}{2}$ particle connected to a local environment, and the dissipation is modeled using an appropriate Lindbladian. The system Hamiltonian in the lab frame is:  
\begin{equation}
\mathcal{H}_{\circ} = \frac{\omega_{\circ}}{2} \sigma_z,
\end{equation}

and an on-resonance external drive along the x-direction is applied: 
\begin{eqnarray}
H_{\text{Drive}} = \frac{\omega_1}{2} \sigma_x
\end{eqnarray}

where, $\sigma_z$ ($\sigma_x$)  is the $z$($x$)-component of Pauli matrices, $\omega_{\circ}$ is the Zeeman frequency of the system and $\omega_{1}$ is the strength of the external drive.

The evolution of density matrix using FRQME (Eq. \ref{frqme_lab}) can be written as: 
\begin{eqnarray} \label{TM_eq}
\dot{\rho}_s &=& - i \frac{\omega_1}{2} [\sigma_x, \, \rho_{\rm s}] - \frac{\omega_1^2 \tau_{c}}{4} [\sigma_x, \, [\sigma_x, \, \rho_{\rm s}]] \nn \\ 
&+& 2\,\omega^2_{\rm SL} \tau_c  \Big(P_1\; \mathcal{D}[\sigma_{+}](\rho_{\rm s}) + P_2 \; \mathcal{D}[\sigma_{-}](\rho_{\rm s}) \Big)
\end{eqnarray}
where, $\omega_{\rm SL}$ is the strength of coupling between system and local environment, $P_{1,2} = e^{\mp \beta \hbar \omega_{\circ}/2}/ \mathcal{Z}$, $\hbar$ is reduced Planck's constant, $1/\beta = K_B T$ is the inverse temperature of the bath, $K_B$ is the Boltzmann's constant, $T$ is the temperature of the bath, $\mathcal{Z}$ is the partition function of the modeled bath, and $\mathcal{D}[L](\rho) =  L \, \rho \, L^{\dagger} - \frac{1}{2} \, L^{\dagger} L  \, \rho - \frac{1}{2} \, \rho \, L^{\dagger} L $ is the Lindbladian dissipator.

We find the equation of evolution of the Magnetization using Eq. \ref{TM_eq} as: 
\begin{eqnarray} \label{Magnetization_Eq}
\dot{M}_{\alpha}(t) = \text{Tr} \left(\frac{\sigma_{\alpha}}{2} \,  \dot{\rho}_s \right) ; \quad \alpha \in \{x, y, z\}
	\end{eqnarray}

Simplifying Eq. \ref{Magnetization_Eq}, we find the following equations for the time evolution of Magnetization: 
\begin{eqnarray} \label{M_eq}
\dot{M}_x(t) &=& - (\Gamma_{\parallel} - \Gamma_{\perp}) \, M_x(t) \\
	\dot{M}_y(t) &=& - \omega_{1} \, M_z(t) - \Gamma_{\perp} M_y(t) \\
	\dot{M}_z(t) &=& \omega_{1} \, M_y(t) - \Gamma_{\parallel} M_z(t) + M_{\circ} \, \omega^2_{\rm SL} \tau_c
	\end{eqnarray}
where, $M_{\circ} = P_1 - P_2$, $\Gamma_{\perp} =  \Gy$, and $\Gamma_{\parallel} \;=\; \Gz$

We note that the transverse and longitudinal rates are \emph{not} equal. The Lindblad dissipator of Eq. \ref{TM_eq} is isotropic in the $x$--$y$ plane, so it damps the coherences $M_{x,y}$ at the rate $\omega^2_{\rm SL}\tau_c$ (half the sum of the up and down rates, $\tfrac{1}{2}(\gamma_{\uparrow}+\gamma_{\downarrow})$) while damping $M_z$ at $2\,\omega^2_{\rm SL}\tau_c = \gamma_{\uparrow}+\gamma_{\downarrow}$. The DID term $-\tfrac{1}{4}\omega_1^2 \tau_c [\sigma_x,[\sigma_x, \cdot]]$ is a pure dephasing generator about the drive axis; it therefore contributes $\omega_1^2 \tau_c$ to both $\Gamma_{\perp}$ (for $M_y$) and $\Gamma_{\parallel}$ (for $M_z$), but leaves $M_x$ untouched.

Since $\Gamma_{\parallel} - \Gamma_{\perp} = \omega^2_{\rm SL}\tau_c$ is of order $\omega_{\rm SL}\chi$ while $\omega_1 \sim \omega_{\rm SL}$ near the optimum, we have $|\Gamma_{\parallel}-\Gamma_{\perp}| \ll \omega_1$, and the transverse--longitudinal splitting may be absorbed into a single mean rate
\begin{eqnarray} \label{Gamma_bar}
\Gamma \;\equiv\; \frac{\Gamma_{\perp} + \Gamma_{\parallel}}{2} \;=\; \Gbar
\end{eqnarray}

with corrections of relative order $(\Gamma_{\parallel}-\Gamma_{\perp})^2/\omega_1^2 = \mathcal{O}(\chi^2)$. With this understanding, we solve Eq. \ref{M_eq} to first order in $\tau_c$ assuming $M_x(0) = 0, M_y(0) = 0, M_z(0) = M_{\circ}$ to get:
\begin{eqnarray}
			M_x(t) &=& 0 \label{Mx}\\
			M_y(t) &=& - M_{\circ} e^{-\Gamma t} \sin(\omega_{1}\, t) + \frac{M_{\circ} \omega_{\rm SL}^2 \tau_{c}}{\omega_{1}} \Big(e^{-\Gamma t} \cos(\omega_{1}\, t) - 1 \Big)+ \mathcal{O}(\tau_{c}^2) \label{My} \\
			M_z(t) &=& M_{\circ} e^{-\Gamma t} \Big(\cos(\omega_{1}\, t) - \frac{\omega_{\rm SL}^2 \tau_{c}}{2 \omega_{1}} \sin(\omega_{1}\, t) \Big) + \mathcal{O}(\tau_{c}^2) \label{Mz}
		\end{eqnarray}

We are interested in finding $\omega_{1}$ that takes the system from $\vert 0 \rangle$ to $\vert 1 \rangle$, which is effectively the action of a $\pi-$pulse. $\langle 1 \vert M_z \vert 1 \rangle$ being the lowest on the Bloch sphere, we minimize $M_z$ with respect to time by taking time derivative of $M_z$ (Eq. \ref{Mz}) and equate to $0$ to get: 

\begin{eqnarray} \label{T_min_MZ}
	\dfrac{d M_z(t)}{dt} = 0 \implies \omega_{1}\, t^{\star} = \pi - \tan^{-1} \left(\frac{\omega_{1}^2 + 2 \omega_{\rm SL}^2}{\omega_{1}} \, \tau_{c}\right)
\end{eqnarray}

When there is no dissipation \textit{i.e.} $\tau_{c}=0$ then we get $\omega_{1} t = \pi $. This means the system reaches $M_z = - M_{\circ}$, which is the minimum value of $M_z$ in the ideal (unitary) case. For $\tau_c \neq 0$, the optimal pulse angle is slightly \emph{less} than $\pi$, the deficit being first order in $\tau_c$.

\subsection{Optimal drive amplitude and optimal time} \label{SSSec: optimum}

Equation \ref{T_min_MZ} determines the optimal pulse \emph{angle}, but not the optimal \emph{speed}. To exhibit the latter, which is the central claim of this work, we must ask, at fixed initial and target states, what drive amplitude $\omega_1$ brings the system closest to $\vert 1 \rangle$. Evaluating Eq. \ref{Mz} at the inversion time $t^{\star} \simeq \pi/\omega_1$ of Eq. \ref{T_min_MZ}, and retaining the leading behaviour, the attainable inversion is
\begin{eqnarray} \label{Mz_peak}
	\left\vert \frac{M_z(t^{\star})}{M_{\circ}} \right\vert & \simeq& e^{-\Gamma t^{\star}} \;=\; \exp\left[- \pi \tau_c \left( \omega_1 + \frac{3\,\omega^2_{\rm SL}}{2\,\omega_1} \right)\right]
\end{eqnarray}

up to corrections of $\mathcal{O}(\tau_c^2)$. The exponent displays, in closed form, precisely the competition described in the Introduction: the term linear in $\omega_1$ is DID, which penalizes \emph{fast} evolution, while the term $\propto 1/\omega_1$ is environmental relaxation accumulated over the pulse duration $t^{\star} = \pi/\omega_1$, which penalizes \emph{slow} evolution. Neither term alone possesses a stationary point; their sum does. Minimizing the exponent with respect to $\omega_1$ gives
\begin{eqnarray} \label{omega_opt}
	\omega_1^{\star} &=& \sqrt{\frac{3}{2}}\;\omega_{\rm SL}, \qquad
	T_{\rm DID} \;=\; \frac{\pi}{\omega_1^{\star}} \;=\; \frac{\pi}{\sqrt{3/2}\;\omega_{\rm SL}}
\end{eqnarray}

and a corresponding ceiling on the attainable inversion,
\begin{eqnarray} \label{fidelity_ceiling}
	\left\vert \frac{M_z(T_{\rm DID})}{M_{\circ}} \right\vert_{\rm max} &=& \exp\left[- \pi \sqrt{6} \; \chi \right], \qquad \chi \equiv \omega_{\rm SL} \tau_c
\end{eqnarray}

Equations \ref{omega_opt} and \ref{fidelity_ceiling} are the analytical core of this
paper, and we draw three conclusions from them. First, there exists a \emph{finite},
non-zero optimal drive amplitude and hence a finite optimal evolution time; the system
cannot be driven arbitrarily fast to advantage, even with unbounded control power. Second,
the optimum is set by the system--environment coupling alone: $\omega_1^{\star}$ is of the
order of $\omega_{\rm SL}$, independent of $\tau_c$. Third,
the maximum attainable fidelity is capped by the single dimensionless parameter $\chi$,
and approaches unity only as $\chi \to 0$; no control protocol whatsoever can beat this
ceiling within the toy model. We emphasize that if DID is discarded (i.e., \ if the
$\omega_1^2\tau_c$ term is dropped from $\Gamma$, as it would be in a standard Lindblad or
Redfield treatment in which the drive enters only to first order), the exponent of Eq.
\ref{Mz_peak} becomes monotonically decreasing in $\omega_1$, the stationary point
disappears, and one recovers the conventional conclusion that faster is always better. The
optimum is thus a genuine consequence of DID and not an artifact of the analysis.

In scaled units ($\varsigma = \omega_{\rm SL} t$) the
prediction of Eq. \ref{omega_opt} reads $T_{\rm DID} \approx 2.6$, with $u^{\star} \approx
1.2$. This is a sharp, parameter-free prediction against which the numerically optimized
protocols of Sec. \ref{Sec: results} can be, and are, compared. We anticipate that the
GRAPE solutions will find a \emph{shorter} optimal time than Eq. \ref{omega_opt}, because
Eq. \ref{Mz_peak} is restricted to a constant-amplitude, on-resonance, single-quadrature
drive, whereas GRAPE is free to shape both quadratures; the toy model therefore provides
an upper bound $T_{\text{DID}}$.

\subsection{Dynamical Equation} \label{SSSec: dynamics} 

Now, to do a rigorous analysis, we consider a more general external drive Hamiltonian applied to the system, in the lab frame, as given below: 
\begin{equation} \label{drive} 
	\mathcal{H}_{\rm dr}(t) = \Big{(} \text{u}_1(t) \sigma_x + \text{u}_2(t)  \sigma_y \Big{)} \cos( \omega t)
\end{equation}

We note that the drive amplitudes, $\text{u}_{1,2}$, are time dependent and $\omega$ is the frequency at which the drive is applied. $\sigma_x$ ($\sigma_y$) is the x (y)-component of Pauli matrix.

The following transformation operator is used to transform the Hamiltonian from the lab frame to the interaction representation of the system Hamiltonian: 
\begin{equation}
	U = e^{i \mathcal{H}_{\circ} t}
\end{equation}

The drive Hamiltonian in the interaction representation takes the following form: 
\begin{eqnarray} \label{Drive IP}
	{H}_{\rm dr}(t) &=& \alpha(t) \; \sigma_{+} \; \Big(  e^{i \Delta_{+} t} + e^{-i \Delta_{-} t} \Big{)}
	+ \alpha^*(t) \; \sigma_{-} \; \Big( e^{-i \Delta_{+} t} +  e^{i \Delta_{-} t} \Big{)} 
\end{eqnarray}

where, $\alpha(t) = \frac{1}{2}\left(\text{u}_1(t) - i\text{u}_2(t)\right) $, $\Delta_{\pm} = \omega \pm \omega_{\circ}$ is the frequency of the counter (co)-rotating component in the rotating frame and $\sigma_{\pm} = \frac{1}{2}\left(\sigma_x \pm i \sigma_y\right)$.

We use the Hamiltonian given in Eq. \ref{Drive IP} and FRQME (Eq. \ref{frqme_lab}) to calculate the complete dynamical equation: 

\begin{eqnarray} \label{Dynamical_Equation}
	\dot{\rho}_s = -i [H_{\rm dr}(t), \; \rho_s]^{\text{sec}}  + \mathcal{D}_{\rm dr}[\rho_{\rm s}] + \mathcal{D}_{\rm SL}[\rho_{\rm s}],
\end{eqnarray}

The first term gives the first-order contribution of the external drive, describing the unitary contribution as given below: 
\begin{eqnarray}
	-i [H_{\rm dr}(t), \; \rho_s]^{\text{sec}} = &-& i\; \alpha(t) \; [\sigma_{+},  \; \rho_{\rm s}] \; e^{-i \Delta_{-} t}
	- i \; \alpha^*(t) \; [\sigma_{-},  \; \rho_{\rm s}] \; e^{i \Delta_{-} t} 
\end{eqnarray}

The second term gives the second-order contribution of the external drive (DID), which is calculated as follows: 
\begin{eqnarray} \label{DID_here}
	\mathcal{D}_{\rm dr}[\rho_{\rm s}] = - \int_{0}^{\infty} d \tau \; e^{-\frac{\tau}{\tau_c}}  \;  [H_{\rm dr}(t), \; [H_{\rm dr}(t - \tau), \; \rho_{\rm s}]]^{\text{sec}} 
\end{eqnarray}

Keeping only the secular terms and doing the necessary integration, we get the final form of DID as: 
\begin{eqnarray} \label{DID_final}
	\mathcal{D}_{\rm dr}[\rho_{\rm s}]  &=& 2 \, \vert \alpha(t) \vert ^2   \Big( \text{J}_\text{R}(\Delta_{+}) + \text{J}_\text{R}(\Delta_{-})  \Big) \Big{(} \mathcal{D}[\sigma_{+}](\rho_{\rm s}) +  \mathcal{D}[\sigma_{-}](\rho_{\rm s})  \Big{)} \nn \\
	&-& \text{J}_\text{R}(\Delta_{-}) \Big{(}    \alpha(t)^2  \; e^{- 2i \Delta_{-} t} \, [\sigma_{+}, [\sigma_{+}, \; \rho_{\rm s}]]+ h.c. \Big{)} \nn \\ 
	&-& i \, \vert \alpha(t) \vert ^2 \, \Big(\text{J}_\text{I}(\Delta_{+})  - \text{J}_\text{I}(\Delta_{-}) \Big) [\sigma_{+} \sigma_{-} - \sigma_{-} \sigma_{+}, \; \rho_{\rm s}] \nn \\ 
	&-& \Big(i \alpha(t)^2 \, \text{J}_\text{I}(\Delta_{-}) [\sigma_{+}, [\sigma_{+}, \; \rho_{\rm s}]] \; e^{- 2i \Delta_{-} t} + h.c.\Big)
\end{eqnarray}

where the spectral density is broken into real part, $\text{J}_\text{R}$ and imaginary part, $\text{J}_\text{I}$: 
\begin{eqnarray}
	\text{J}(x) &=& \int_{0}^{\infty} d \tau \; e^{-\frac{\tau}{\tau_c}}  \;  e^{i x \tau}
	= \frac{\tau_c}{1+(x \tau_c)^2} + i \frac{x \tau_c^2}{1+(x \tau_c)^2} = \text{J}_\text{R}(x) + i \text{J}_\text{I}(x)
\end{eqnarray}

There are two types of terms in Eq. \ref{DID_final}. The terms with $\text{J}_\text{R}$ are the dissipative terms, and the terms with $\text{J}_\text{I}$ are the shift terms that shift the energy states of the system.

The third term of Eq. \ref{Dynamical_Equation} gives the dissipation due to the environment, which is considered to be in the standard Lindblad form:
\begin{eqnarray} \label{Env_Diss}
	\mathcal{D}_{\rm SL}[\rho_{\rm s}] &=& 2\,\omega^2_{\rm SL} \tau_c  \Big(P_1\; \mathcal{D}[\sigma_{+}](\rho_{\rm s}) + P_2\; \mathcal{D}[\sigma_{-}](\rho_{\rm s}) \Big)
\end{eqnarray}

We scale the entire dynamical equation with $\omega_{\rm SL}$, to get: 
\begin{eqnarray} \label{final rho}
	\dot{\rho}_{s} &=& - i \, \text{u}(\varsigma) \; [\sigma_{+},  \rho_{\rm s}]e^{-i \delta_{-} \varsigma} -  i \, \text{u}^{*}(\varsigma) \; [\sigma_{-},  \rho_{\rm s}]e^{i \delta_{-} \varsigma}  \nn \\ 
	&-& i \, \vert \text{u}(\varsigma) \vert ^2 \, \Big(\text{J}_\text{I}(\delta_{+})  - \text{J}_\text{I}(\delta_{-}) \Big) [\sigma_{+} \sigma_{-} - \sigma_{-} \sigma_{+}, \rho_{\rm s}] \nn \\ 
	&-& \Big(i \text{u}(\varsigma)^2 \, \text{J}_\text{I}(\delta_{-}) [\sigma_{+}, [\sigma_{+}, \rho_{\rm s}]] \; e^{- 2i \delta_{-} \varsigma} + h.c.\Big) \nn \\
	&+ & \left( \Gamma_{\rm dr} + 2\chi P_1 \right) \mathcal{D}[\sigma_+](\rho_{\rm s}) +  \left( \Gamma_{\rm dr} + 2\chi P_2 \right) \mathcal{D}[\sigma_-] (\rho_{\rm s})  \nn \\
	&-& J_\text{R}(\delta_{-}) \Big{(}     \text{u}(\varsigma)^2  \; e^{- 2i \delta_{-} \varsigma} \, [\sigma_{+}, [\sigma_{+}, \rho_{\rm s}]]+ h.c. \Big{)}
\end{eqnarray}

where, $\varsigma = \omega_{\rm SL}t$; $\text{u}(\varsigma) = \frac{\alpha(\varsigma)}{\omega_{\rm SL}} $; $\delta_{+,-} = \frac{\Delta_{+,-}}{\omega_{\rm SL}}$; 
$\Gamma_{\rm dr} = 2 \, \vert \text{u}(\varsigma) \vert^2 \Big(J_\text{R}(\delta_{+})  + J_\text{R}(\delta_{-}) \Big)$ is the drive-induced dissipation (DID) rate;
$\chi = \omega_{\rm SL}  \tau_c$ is the \emph{scaled} (dimensionless) environmental correlation time.

Finally, we note that the double-commutator terms in Eq. \ref{DID_final} involve
non-Hermitian operators and are therefore not manifestly of Lindblad form. On resonance
($\delta_- = 0$) they collapse to $-\tfrac{1}{4}\omega_1^2 \tau_c
[\sigma_x,[\sigma_x,\rho_s]]$, which is a bona fide dephasing generator and is completely
positive.


\section{Implementation of GRAPE algorithm} \label{Sec: GRAPE}

Gradient ascent pulse engineering (GRAPE) is used to engineer the pulse profile that will give us the shortest possible trajectory to go from initial state $\vert \psi_i \rangle$ to as close as possible to target state $ \vert \psi_t \rangle$ in a specified number of steps \cite{khaneja2005optimal, schulte2011optimal}. The time of evolution and hence the drive Hamiltonian is discretized into $N$ steps. The $j^{th}$ time step is between $t_{j-1}$ to $t_j$. It is assumed that the drive Hamiltonian for a specific time step remains constant. The Hamiltonian at $j^{th}$ time step is:
\begin{equation}
	H(j) = H_{\circ} + \sum_{k = 1}^m \text{u}_k(j) H_k
\end{equation}

\noindent where, $H_{\circ}$ is the bare Hamiltonian, $\text{u}_k(j)$ is amplitude of the $k^{th}$ control field at $j^{th}$ time step, and $H_k$ is operator for the $k^{th}$ control field

The control amplitude is updated at each step in the following way: 
\begin{equation} \label{update rule}
	u_{k}(j) \rightarrow u_{k}(j) + \frac{\delta \phi_{k}}{\delta u_{k}(j)}
\end{equation}

Here, $\phi_{k} = h_k (1 - \text{F}(\rho_{\rm T}, \rho_{\rm f}))$ is the performance
index, where, $\text{F}(\rho_{\rm T}, \rho_{\rm f})$ is the fidelity between $\rho_{\rm
	T}$ and $\rho_{\rm f}$. $h_k$ is a dynamically chosen scalar that determines how much to
Update the parameters along the search direction in each iteration. $ \frac{\delta
	\phi_{k}}{\delta u_{k}(j)}$ gives the gradient of performance index with respect to
control amplitude for $k^{th}$ control at $j^{th}$ time step, $\rho_{\rm T}$ is the
density matrix of the target state and $\rho_{\rm f}$ is the density matrix of the final
state.

We initially guess the control amplitudes of the drive Hamiltonian ($\text{u}_{1, 2}$)
randomly. Then, using the dynamical equation (Eq. \ref{final rho}), we find the final
density matrix ($\rho_{\rm f}$) and calculate the fidelity with respect to the target
state ($\rho_{\rm T}$). The gradient is calculated in the parameter space of the control
amplitudes. The next step is taken in the direction of the minimum gradient. Accordingly,
the control amplitude is updated using Eq. \ref{update rule}, which is then used as the
initial guess for the next iteration. The algorithm repeats the process until it
converges. At every step, the pulse profile is modified, and finally, we get an optimized
pulse profile.

We stress one methodological point, because the central claim of this work depends on
it. \emph{No upper bound is imposed on the control amplitudes} $\text{u}_{1,2}(j)$ during
the optimization: the algorithm is free to raise the drive power without limit in order to
shorten the evolution. Had the amplitudes been capped, an apparent optimal time would
arise trivially, as an artifact of the cap; below some duration, the capped drive simply
cannot complete the rotation. The finite $T_{\rm QOC}$
reported in Sec. \ref{Sec: results} is therefore \emph{not} an effect arising from
imposition of an ad hoc upper limit. It
arises because the DID contribution $\Gamma_{\rm dr} \propto
\vert \text{u}(\varsigma)\vert^2$ in Eq. \ref{final rho} penalizes the very power that
shortening the pulse requires, exactly as in Eq. \ref{Mz_peak}. To confirm this, we have
repeated every optimization with the DID terms of Eq. \ref{DID_final} removed by hand,
retaining only the environmental Lindblad dissipator of Eq. \ref{Env_Diss}; the results
are reported in Sec. \ref{Sec: results} and are shown in Fig.
\ref{fig:Optimality_Bloch_Sphere}.

The number of time steps was fixed at $N$ and the fidelity and $T_{\rm QOC}$ were checked
for convergence with respect to $N$. 


\section{Numerical Results} \label{Sec: results}

\begin{figure*}[t]
	\centering
	\includegraphics[scale=0.7]{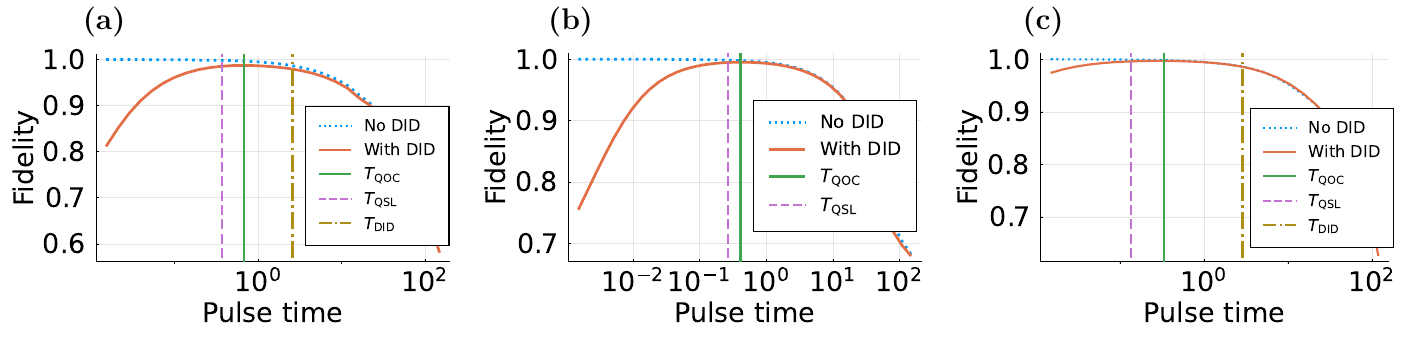}
	\caption{The figure shows the variation in fidelity with the change in pulse time with and without DID for the evolution of the qubit from (a) +Z to -Z, (b) +Z to +X, (c) +S to +R. States are as defined in Eq. \ref{Z_X} and Eq. \ref{S_R}. The parameters used for
		simulation are $\frac{\omega_{\circ}}{\omega_{\rm SL}} = 5
		\times 10^{3}$,  $\delta_{-} = 0$, $\delta_{+} = \delta_{-} + \frac{2
			\omega_{\circ}}{\omega_{\rm SL}}$, $T = 37$ mK, $\chi = 5
		\times 10^{-3}$. Here, $T_{\rm QOC}$ is the minimum time
		corresponding to maximum fidelity found using quantum optimal control when DID is included, $T_{\rm QSL}$ is the QSL time calculated using Eq. \ref{T qsl} and $T_{\rm DID}$ is the time calculated using the toy model given in Sec. \ref{Sec: Model}. The control
		amplitudes are unconstrained; see Sec. \ref{Sec: GRAPE}. }
	\label{fig:Optimality_Bloch_Sphere}
\end{figure*}

To find the shortest time of evolution that evolves the system closest to the target state, we solved Eq. \ref{final rho} numerically by varying the step size ($\Delta t$) of the applied external pulse, keeping the number of pulses ($N$) constant. The total time of application of the pulse $T_{\rm pulse} = N \Delta t$ and the corresponding fidelity were plotted on the X-axis and Y-axis of Fig. \ref{fig:Optimality_Bloch_Sphere}, respectively. The evolution of the system from different initial to target states is shown in Fig. \ref{fig:Optimality_Bloch_Sphere}(a,b,c). The same behavior was observed when $\Delta t$ was kept constant and $N$ was varied. We see that there exists an optimal value $T_{\rm QOC}$, corresponding to which we get maximum fidelity. Maximum fidelity implies the system is very close to the desired target state. The trajectory reaches closer to the target state, but not exactly on the target state, because the system suffers from losses due to environmental dissipation and DID. The standard states used to plot all the figures are: 
\begin{eqnarray} \label{Z_X}
	\vert  +Z \rangle = \vert 0 \rangle, \hspace{0.2cm} \vert -Z \rangle = \vert 1 \rangle, \hspace{0.2cm} \vert \pm X \rangle = \frac{1}{\sqrt{2}}(\vert 0 \rangle \pm \vert 1 \rangle)
\end{eqnarray}

The method works for the evolution of the system from any given initial state to a target state. To demonstrate that, we start with an arbitrary state $\vert + S \rangle$ and evolve to a target state $\vert + R \rangle$:
			\begin{eqnarray} \label{S_R}
				\vert + S \rangle &=& \cos \left(\frac{\pi}{8} \right) \vert 0 \rangle + \;  \sin \left(\frac{\pi}{8} \right) \vert 1 \rangle \nn \\
				\vert + R \rangle &=& \cos \left(\frac{\pi}{8} \right) \vert 0 \rangle + i \sin \left(\frac{\pi}{8} \right) \vert 1 \rangle 
			\end{eqnarray}
The time evolution is shown in Fig. \ref{fig:Optimality_Bloch_Sphere}(c). This shows that the method also works when the initial and target states are not on the transverse plane of the Bloch Sphere. 

\begin{figure*}[t]
	\centering
	\includegraphics[scale=1.1]{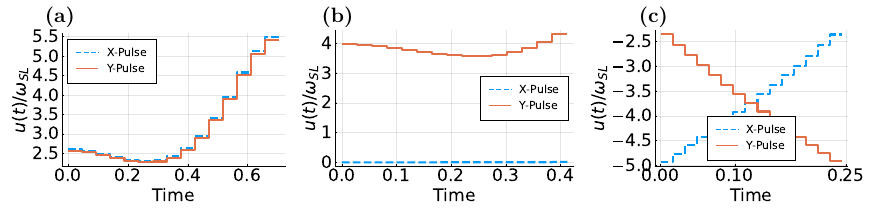}
	\caption{The figure shows an engineered pulse profile to drive the qubit in optimal time ($T_{\rm QOC}$) given in Fig. \ref{fig:Optimality_Bloch_Sphere}(a-c) from (a) +Z to -Z (b) +Z to +X (c) +S to +R. States are as defined in Eq. \ref{Z_X} and Eq. \ref{S_R}. The parameters used for simulation are $\frac{\omega_{\circ}}{\omega_{\rm SL}} = 5 \times 10^{3}$,  $\delta_{-} = 0$, $\delta_{+} = \delta_{-} + \frac{2 \omega_{\circ}}{\omega_{\rm SL}}$, $T = 37$ mK, $\chi = 5 \times 10^{-3}$}
	\label{fig: Pulse_Profile}
\end{figure*}

The use of the GRAPE algorithm allows us to find the optimal pulse profile that evolves the system from a given initial state to a target state as close as possible in the minimum possible time. Fig. \ref{fig: Pulse_Profile}(a,b,c) shows the optimal pulse profile for evolving the system from a given initial state to any target state. When going from +Z to +X, the Y-pulse is nonzero because it drives the system to the target state. On the other hand, for the arbitrary states, both pulses need to drive the system. This pulse combination acts as an effective Z-pulse. We note that when trying to find the optimal pulse profile for the arbitrary states, giving the optimal pulse profile for the +Z to +X transition as an initial guess helps in better convergence.

\begin{figure*}[t]
	\centering
	\includegraphics[scale=1.2]{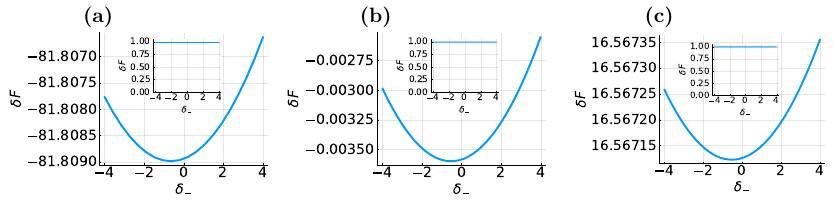}
	\caption{The figure shows variation in fidelity ($\delta F$) as a function of detuning
		($\delta_{-}$) for evolution of the system from (a) +Z to -Z (b) +Z to +X (c) +S to +R.
		Here, $\delta F = \left(F - F_0\right) \times 10^{4}$ is chosen to show the effects of the
		variation in fidelity ($F$) with $F_0 = 0.99525$. The inset shows that for the full range
		of fidelity $[0,1]$ the variation is negligible. States are as defined in Eq. \ref{Z_X}
		and Eq. \ref{S_R}. The parameters used for simulation are
		$\frac{\omega_{\circ}}{\omega_{\rm SL}} = 5 \times 10^{3}$,
		$\delta_{-} = 0$, $\delta_{+} = \delta_{-} + \frac{2
			\omega_{\circ}}{\omega_{\rm SL}}$, $T = 37$ mK, $\chi = 5
		\times 10^{-3}$. (Note that $\delta_{-}$ is the abscissa of this figure and is therefore not a fixed parameter of it; the listing above is retained only for the
		remaining parameters.) }
	\label{fig: Detuning}
\end{figure*}

The analysis so far was performed by setting the drive at zero-detuning frequency
($\delta_- = 0$). To check the robustness of the method, we varied the detuning from $-4$
to $4$ (in scaled units) as shown in Fig. \ref{fig: Detuning}(a,b,c) for different initial
states and different target states. The fidelity varies by only $\sim 2\times 10^{-7}$ in absolute terms
across this range, i.e., it is almost unaffected by the variation in detuning. 

\begin{figure*}[t]
	\centering
	\includegraphics[scale=0.6]{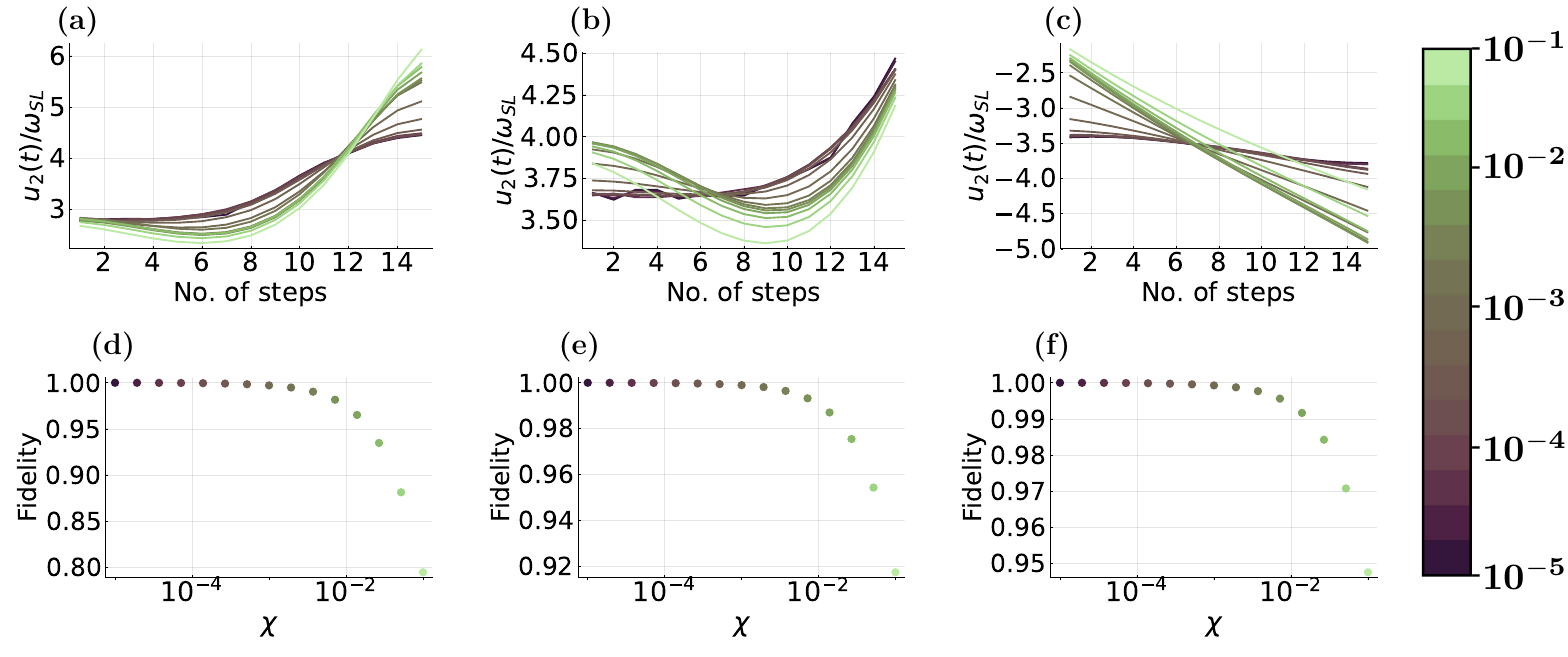}
	\caption{The figure shows variation in optimal pulse profile for Y-quadrature ($\text{u}_2(t)/\omega_{\rm SL}$) when the environmental correlation time $\chi$ was varied from $10^{-5} \; \rm{to} \; 10^{-1}$ (purple to teal, color online) for evolution of the system from (a) +Z to -Z (b) +Z to +X (c) +S to +R. The corresponding fidelity variation with $\chi$ is shown in (d), (e), and (f), respectively. States are as defined in Eq. \ref{Z_X} and Eq. \ref{S_R}. The parameters used for simulation are $\frac{\omega_{\circ}}{\omega_{\rm SL}} = 5 \times 10^{3}$,  $\delta_{-} = 0$, $\delta_{+} = \delta_{-} + \frac{2 \omega_{\circ}}{\omega_{\rm SL}}$, $T = 37$ mK. The low values ($\chi \sim 10^{-5}$) in the figure appear in deep purple (dark gray), which transition to bright teal (light gray) at high values ($\chi \sim 10^{-1}.$)}
	\label{fig: Band}
\end{figure*}

Another way to check the robustness is to vary the environment correlation time ($\chi$) and
see how the optimal profile changes. The value of $\chi$ was varied from $10^{-5} \;
\rm{to} \; 10^{-1}$ (in scaled units) as shown in Fig. \ref{fig: Band}(a,b,c) for
different initial and final state. While their corresponding fidelity variation is plotted
in Fig. \ref{fig: Band}(d,e,f). We observe that the corresponding fidelity is $\approx1$
for  $\chi = 10^{-5} \; \rm{to} \; 10^{-2}$. This corroborates our claim that the method
is robust under a wide range of environmental correlation times.

We take the hint from Fig. \ref{fig: Band} and try to find the optimal parameter regime where we are certain to get maximum possible fidelity. We vary the time of evolution by varying $\Delta t$ and $\chi$ and show the variation in the contour plots of Fig. \ref{fig: contour}(a,b,c) for different initial and final states. The corresponding fidelity is shown in the color bar.  We observe that a lower $\chi$ gives higher fidelity irrespective of the total time of evolution. As the value of $\chi$ increases, the fidelity begins to decrease. This is expected as the increase in $\chi$ corresponds to an increase in dissipation, which prevents the system from reaching the target state.

\subsection{The optimum is a consequence of DID} \label{SSec: noDID}

The non-monotonicity of Fig. \ref{fig:Optimality_Bloch_Sphere}(a,b,c) is the central
numerical observation of this work. To establish that this indeed originates  DID rather
than in some feature of the optimization, we repeated the entire calculation with the
drive-induced dissipator $\mathcal{D}_{\rm dr}[\rho_s]$ of
Eq. \ref{DID_final} set to zero, retaining the environmental Lindblad term of Eq.
\ref{Env_Diss} and leaving the control amplitudes unconstrained as before. In that case,
the fidelity rises \emph{monotonically} as $T_{\rm pulse} \to 0$ and saturates at unity:
with no penalty on drive power, GRAPE simply raises the amplitude without limit and
completes the transfer before the environment can act. We do not find any optimum time. Restoring
$\mathcal{D}_{\rm dr}$ restores the optimum. This
clearly demonstrates that the two-sided window described in the Introduction
is a physical consequence of DID and not an artifact of limiting the drive power, of the
discretization, or of the choice of cost function.

\subsection{Comparison with the analytical bound} \label{SSec: TQSL}

Table \ref{tab: TQSL} collects, for each of the three transitions, the numerically optimal
time $T_{\text{QOC}}$, the maximum attainable fidelity
$F_{\rm max}$, the analytical bound $T_{\rm QSL}$ evaluated from Eq. \ref{T qsl} using the
GRAPE pulse and the $T_{\rm DID}$.

First, we observe that the inequality $T_{\rm QSL} \le T_{\rm QOC} \le T_{\rm DID}$ holds
true for each of the investigated cases.  For the investigated cases, the QOC optimum
therefore lies between the QSL lower bound and the constant-drive optimal timescale $T_{\rm
DID}$.

\begin{table}[b]
	\caption{Numerically optimal evolution time $T_{\text{
				QOC}}$ (scaled units, $\varsigma = \omega_{\rm SL} t$)
		compared with the analytical bound $T_{\rm QSL}$ of Eq. \ref{T qsl} evaluated at the
		achieved Bures angle. Parameters as in Fig. \ref{fig:Optimality_Bloch_Sphere}.}
	\label{tab: TQSL}
	\begin{ruledtabular}
		\begin{tabular}{p{3cm} p{2.5cm} p{2.5cm} p{2.5cm} p{2.5cm}}
			Transition & $F_{\rm max}$ & $T_{\rm QSL}$  & $T_{\rm QOC}$ & $T_{\rm DID}$ \\
			\hline
			$+Z \to -Z$ & 0.99 & 0.370 & 0.68 & 2.6 \\
			$+Z \to +X$ & 1.00 & 0.267 & 0.41 &  $\infty$ \\
			$+S \to +R$ & 1.00 & 0.136  & 0.34 & 2.87 \\
		\end{tabular}
	\end{ruledtabular}
\end{table}

Second, and importantly, the bound of Eq. \ref{T qsl} taken \emph{by itself} exhibits no
optimum: every term in $\lambda_1$ increases with drive amplitude, so $T_{\rm QSL}$
decreases monotonically as the drive is strengthened and can be made arbitrarily small.
The bound therefore constrains the lower edge $T_{\min}$ of the feasibility window but
says nothing about the upper edge. The optimum reported here arises entirely from the fact
that the attainable $\theta$ cannot be made small at large drive amplitude, i.e.\ from the
fidelity ceiling of Eq. \ref{fidelity_ceiling}. The bound and the optimum are
complementary statements, and neither subsumes the other.


\begin{figure*}[t]
	\centering
	\includegraphics[scale=1]{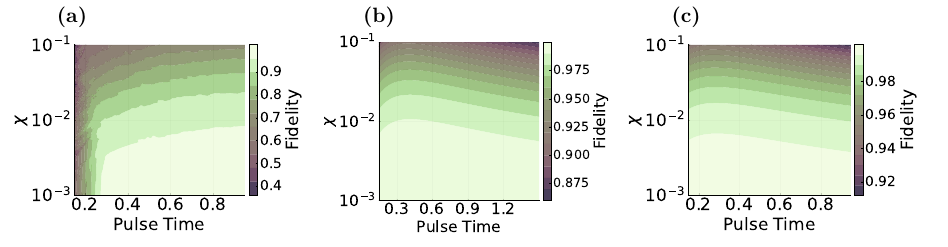}
	\caption{The figure shows a contour plot showing the change in fidelity as we change the environmental correlation time ($\chi$) and time of evolution. (a) from +Z state to -Z state (b) from +Z state to +X state. (c)  from +S state to +R state. States are as defined in Eq. \ref{Z_X} and Eq. \ref{S_R}. The parameters used for simulation are $\frac{\omega_{\circ}}{\omega_{\rm SL}} = 5 \times 10^{3}$,  $\delta_{-} = 0$, $\delta_{+} = \delta_{-} + \frac{2 \omega_{\circ}}{\omega_{\rm SL}}$, $T = 37$ mK.  The low values of fidelity in the figure appear in deep purple (dark gray), which transition to bright teal (light gray) at high values of fidelity.}
	\label{fig: contour}
\end{figure*}


\section{Discussion} \label{Sec: discussion}

In practical scenarios, a quantum system can never reach its target state exactly due to
dissipation and noise, which are unavoidable. As a result, the system's final density
matrix always remains within the Bloch Sphere. To overcome these inherent limitations, we
need strategies to maximize fidelity in realistic systems. In this work, we have revisited
the method to evolve a quantum system from an initial state to a final state in the
minimum possible time with maximum possible fidelity by including both the environmental
dissipation as well as dissipation due to external drive. This makes our analysis more
realistic and closer to practical situations. 

Additionally, we observe in Fig. \ref{fig:Optimality_Bloch_Sphere}(a,b,c) that there
exists an optimal pulse time corresponding to which we get maximum fidelity, beyond which
it decreases. This observation can be explained by considering the effects of both the
sources of dissipation. If we evolve the system fast enough, we need to apply stronger
pulses, which result in DID. This reduces the fidelity for shorter pulse times. On the
other hand, if we let the system evolve slowly, the system decoheres because of
environmental dissipation. This reduces the fidelity for longer pulse times. As a result
of competition between these effects, we get an optimal pulse time which gives maximum
fidelity.    

However, the competing effects of environmental dissipation and DID make it difficult to
achieve high-fidelity quantum gate operations. Lower fidelity indicates that the pulse
profile is imperfect. To overcome this, typically faster or multiple pulses are applied,
which increases DID. Our analysis provides a method to find an optimal parameter regime
which minimizes dissipation while maximizing fidelity by balancing these two detrimental
effects as shown in Fig. \ref{fig: contour}. 

We also calculated the quantum speed limit (QSL) bound for our system with environmental
dissipation and drive-induced dissipation taken into account. The detailed calculation is
shown in Appendix \ref{Sec: Analytical QSL}. The comparison between
$T_{\text{QOC}}$ and $T_{\rm QSL}$ is given quantitatively
in Table \ref{tab: TQSL} and requires careful interpretation. The bound of Eq. \ref{T qsl}
is evaluated at the Bures angle $\theta$ actually achieved by the protocol, and is in that
sense tight by construction; the statement ``our optimal time approaches the QSL bound"
therefore carries less content than it might appear to, and we no longer make it in that
form. The substantive content of the comparison is the \emph{ratio}
$T_{\text{QOC}}/T_{\rm QSL}$, which quantifies the slack
left by GRAPE at the fidelity it attains. It is also worth restating that the bound alone
predicts no optimum whatsoever: as noted in Sec. \ref{SSec: TQSL}, every term of
$\lambda_1$ grows with the drive amplitude, so $T_{\rm QSL} \to 0$ as the drive is
strengthened without limit. The finite optimum is a statement about \emph{attainability},
about the ceiling on $\cos^2\theta$ imposed by Eq. \ref{fidelity_ceiling}, and not about
the bound. Conflating the two would be an error, and we are explicit that the QSL, as a
lower bound on the time to reach a \emph{given} state, is nowhere violated in this work.

Our results also show a relationship between environmental correlation time ($\tau_c$) and
fidelity. We observe that a lower $\tau_c$ gives higher fidelity and vice versa. This can
be intuitively understood as lower $\tau_c$ implies lower dissipation, allowing the system
to reach closer to the target state. The plot shown in Fig. \ref{fig: Band}(d,e,f) between
the scaled (dimensionless) correlation time ($\chi =
\omega_{\rm SL} \tau_c$) and fidelity shows that by
carefully choosing the value of $\chi$, which is related to $\tau_c$, we can improve the
fidelity of the quantum operation. 

We chose the GRAPE algorithm over other optimal control methods, such as Krotov's method
\cite{konnov1999global}, or the chopped random basis (CRAB) method
\cite{caneva2011chopped}, because the time-discretized pulse profile used in GRAPE allows
for evaluating the gradient at each time step efficiently. This is well-suited for
low-dimensional systems such as the 2-level system considered in this work. GRAPE offers
high-fidelity state transfer and also flexibility for generic control problems, as there
is no shape constraint on the pulse profile \cite{machnes2011comparing}. Importantly, we
combined GRAPE with the Limited-memory Broyden–Fletcher–Goldfarb–Shanno (L-BFGS) algorithm
to update $h_k$ at each step. This helps to achieve rapid convergence to high fidelity
near optima. The major advantage of GRAPE over other quantum control algorithms is that it
leads to faster convergence as it employs concurrent-update schemes
\cite{machnes2011comparing}.

However, GRAPE is known to be sensitive to the choice of initial pulse profile. But in the
present setting, we optimized the pulse profile for evolving the system from +Z to +X
using a random initial guess. We observed that consistently the algorithm converges to the
same high fidelity solution across multiple random initializations, consistent with a
trap-free control landscape for this system \cite{rabitz2004quantum}. We repeated this
multiple times with independent random seeds and then used this optimal pulse profile as
an initial guess for all other evolutions. We found this results in robust,
high-fidelity, and complication-free optimization. 

We have used infidelity ($1 - \text{F}$) as our cost parameter to find the optimal pulse
profile that minimizes the evolution time and maximizes accuracy. However, any other cost
function, such as trace distance, entropy, Bures angle, or purity, can be used without
changing the underlying physics. 

Figure \ref{fig: Detuning} shows the variation of the fidelity with detuning. Over the
range $\delta_- \in [-4, 4]$ the fidelity varies by $\sim 2\times 10^{-7}$ in absolute
terms, so the protocol is essentially insensitive to errors in the drive frequency, with
the absolute fidelity limited entirely by dissipation rather than by detuning. However,
in experiments, infidelity grows exponentially with the number of repetitions. For
example, if each gate has fidelity $F$, a circuit of $n$ gates has approximate fidelity
$F^n$. For $F= 0.995$ and $n=100$ gates, $F^{100} \approx 0.61$ which is very poor. So
even a small fine-tuning in fidelity means a lot of improvement when state preparation is
repeated or when an optimal control loop is used. Also, improvement in fidelity implies
moving from a fragile optimum to a stable optimum in the parameter space of time-dependent
drive amplitudes. So this result can guide experimentalists in choosing the right value of
detuning to minimize the errors.


Moreover, the optimal pulse time of quantum operation (in non-scaled units) as given by
our method with $\omega_{\rm SL} = 2\pi \text{ MHz}$
\begin{eqnarray}
	T_{\rm non-scaled} &=& \frac{T_{\rm QOC}}{\omega_{\rm SL}}
\end{eqnarray}
is of the order of $100$ ns for $T_{\rm QOC} \sim
0.5$--$1$ in scaled units, which is much shorter compared to the typical coherence time of
trapped ions \cite{wang2017single}, superconducting qubits
\cite{devoret2013superconducting}, neutral atoms \cite{ebert2015coherence}, quantum dots
\cite{veldhorst2014addressable}, and NV centers in diamonds \cite{doherty2013nitrogen}
which ranges between $ms$ to $\mu s$, and in some cases it exceeds a few seconds. So the
method presented here can be verified experimentally using some realistic quantum systems. 

It is fair to ask whether the optimum reported here is specific to FRQME or whether any
master equation carrying a drive-dependent dissipation rate would produce it. The
qualitative mechanism, power costs coherence, so infinite speed is not free, is not by
itself novel: amplitude-dependent decoherence is well documented in driven superconducting
qubits, Floquet--Born--Markov treatments generate relaxation rates that depend on the
drive amplitude, and relaxation-optimized pulse design in magnetic resonance has long
recognized that finite optimal durations emerge once relaxation is included in the
optimization. We do not claim otherwise. What is new here is (i) that the optimum can be
obtained in \emph{closed form} (Eqs. \ref{omega_opt}--\ref{fidelity_ceiling}), so that the
optimal amplitude, the optimal time, and the fidelity ceiling are all expressed in terms
of the two microscopic parameters $\omega_{\rm SL}$ and
$\tau_c$ alone; (ii) that the analytical QSL bound has been extended to include the
regularized second-order drive contribution, yielding the term $\mathcal{Z}(H_\tau, H)$ of
Eq. \ref{ZDID}, which to our knowledge has not appeared before; and (iii) that the
decomposition of $\lambda_1$ answers, quantitatively, the question of whether the
resulting bound remains a ``time-energy" bound, as we find the MT bound remains the
shortest.

We note in passing that the parameters used throughout correspond to a physically
realistic device: $\omega_{\circ}/2\pi = 5$ GHz, $\tau_c \approx 0.8$ ns, an environmental
relaxation time $T_1 \approx 16\ \mu$s, and $\beta \hbar \omega_{\circ} \approx 6.5$ at $T
= 37$ mK. These are representative of a superconducting transmon, and we state them
explicitly so that the experimental relevance of the optimum is not left implicit.

As quantum optimal control is limited by the computational complexity of the numerical
simulation, the analysis is done for a single 2-level system. Still, it offers interesting
insights that can be useful for performing a more complex many-body system.  


\section{Conclusion}
We report that in an open quantum system, with DID taken into account, there exists an
optimal time of evolution, which allows the system to evolve with maximum fidelity in the
minimum possible time.  This implies that DID introduces an additional restriction on
attainability, so that the conventional QSL lower bound is complemented by a finite-time
window whose characteristics and interior optimum are determined by the competition
between drive-induced and environmental dissipation.  Within a solvable toy model, we
obtained this optimum in closed form, an optimal Rabi amplitude $\omega_1^{\star} =
\sqrt{3/2}\;\omega_{\rm SL}$ and a fidelity ceiling $\exp(-\pi\sqrt{6}\,\chi)$, and we
confirmed numerically, using unconstrained control amplitudes and an explicit DID-free
control calculation, that the optimum vanishes when DID is removed. Our method connects
the quantum optimal control with quantum speed limits for open quantum systems. The method
was shown to be robust against detuning and against variation of the correlation time over
several orders of magnitude. Using this method, the quantum system can be evolved from a
given state to any target state with maximum attainable fidelity, at a time which we
compare quantitatively against the extended QSL bound derived in Appendix \ref{Sec:
Analytical QSL}. The method is general and can be adapted to quantum platforms that can
be represented as a 2-level system, offering an approach for high-fidelity gate
design and state preparation under realistic situations. These findings provide a solid
foundation for advancing practical quantum technologies.

\section{Acknowledgment}
SF would like to thank Arpan Chatterjee, Gourab Das, and Shubhamay Panja for insightful
discussions. SF acknowledges the University Grants Commission (UGC) of the Government of
India for providing financial assistance through a junior research fellowship (JRF) with
reference number 221610075321.

\bibliographystyle{apsrev4-2}
\bibliography{references.bib}

\begin{thebibliography}{52}%
\makeatletter
\providecommand \@ifxundefined [1]{%
 \@ifx{#1\undefined}
}%
\providecommand \@ifnum [1]{%
 \ifnum #1\expandafter \@firstoftwo
 \else \expandafter \@secondoftwo
 \fi
}%
\providecommand \@ifx [1]{%
 \ifx #1\expandafter \@firstoftwo
 \else \expandafter \@secondoftwo
 \fi
}%
\providecommand \natexlab [1]{#1}%
\providecommand \enquote  [1]{``#1''}%
\providecommand \bibnamefont  [1]{#1}%
\providecommand \bibfnamefont [1]{#1}%
\providecommand \citenamefont [1]{#1}%
\providecommand \href@noop [0]{\@secondoftwo}%
\providecommand \href [0]{\begingroup \@sanitize@url \@href}%
\providecommand \@href[1]{\@@startlink{#1}\@@href}%
\providecommand \@@href[1]{\endgroup#1\@@endlink}%
\providecommand \@sanitize@url [0]{\catcode `\\12\catcode `\$12\catcode
  `\&12\catcode `\#12\catcode `\^12\catcode `\_12\catcode `\%12\relax}%
\providecommand \@@startlink[1]{}%
\providecommand \@@endlink[0]{}%
\providecommand \url  [0]{\begingroup\@sanitize@url \@url }%
\providecommand \@url [1]{\endgroup\@href {#1}{\urlprefix }}%
\providecommand \urlprefix  [0]{URL }%
\providecommand \Eprint [0]{\href }%
\providecommand \doibase [0]{https://doi.org/}%
\providecommand \selectlanguage [0]{\@gobble}%
\providecommand \bibinfo  [0]{\@secondoftwo}%
\providecommand \bibfield  [0]{\@secondoftwo}%
\providecommand \translation [1]{[#1]}%
\providecommand \BibitemOpen [0]{}%
\providecommand \bibitemStop [0]{}%
\providecommand \bibitemNoStop [0]{.\EOS\space}%
\providecommand \EOS [0]{\spacefactor3000\relax}%
\providecommand \BibitemShut  [1]{\csname bibitem#1\endcsname}%
\let\auto@bib@innerbib\@empty
\bibitem [{\citenamefont {Deffner}\ and\ \citenamefont
  {Campbell}(2017)}]{deffner2017quantum}%
  \BibitemOpen
  \bibfield  {author} {\bibinfo {author} {\bibfnamefont {S.}~\bibnamefont
  {Deffner}}\ and\ \bibinfo {author} {\bibfnamefont {S.}~\bibnamefont
  {Campbell}},\ }\href@noop {} {\bibfield  {journal} {\bibinfo  {journal}
  {Journal of Physics A: Mathematical and Theoretical}\ }\textbf {\bibinfo
  {volume} {50}},\ \bibinfo {pages} {453001} (\bibinfo {year}
  {2017})}\BibitemShut {NoStop}%
\bibitem [{\citenamefont {Mandelstam}(1945)}]{mandelstam1945uncertainty}%
  \BibitemOpen
  \bibfield  {author} {\bibinfo {author} {\bibfnamefont {L.}~\bibnamefont
  {Mandelstam}},\ }\href@noop {} {\bibfield  {journal} {\bibinfo  {journal} {J.
  Phys.(USSR)}\ }\textbf {\bibinfo {volume} {9}},\ \bibinfo {pages} {249}
  (\bibinfo {year} {1945})}\BibitemShut {NoStop}%
\bibitem [{\citenamefont {Margolus}\ and\ \citenamefont
  {Levitin}(1998)}]{margolus1998maximum}%
  \BibitemOpen
  \bibfield  {author} {\bibinfo {author} {\bibfnamefont {N.}~\bibnamefont
  {Margolus}}\ and\ \bibinfo {author} {\bibfnamefont {L.~B.}\ \bibnamefont
  {Levitin}},\ }\href@noop {} {\bibfield  {journal} {\bibinfo  {journal}
  {Physica D: Nonlinear Phenomena}\ }\textbf {\bibinfo {volume} {120}},\
  \bibinfo {pages} {188} (\bibinfo {year} {1998})}\BibitemShut {NoStop}%
\bibitem [{\citenamefont {Giovannetti}\ \emph {et~al.}(2003)\citenamefont
  {Giovannetti}, \citenamefont {Lloyd},\ and\ \citenamefont
  {Maccone}}]{giovannetti2003quantum}%
  \BibitemOpen
  \bibfield  {author} {\bibinfo {author} {\bibfnamefont {V.}~\bibnamefont
  {Giovannetti}}, \bibinfo {author} {\bibfnamefont {S.}~\bibnamefont {Lloyd}},\
  and\ \bibinfo {author} {\bibfnamefont {L.}~\bibnamefont {Maccone}},\
  }\href@noop {} {\bibfield  {journal} {\bibinfo  {journal} {Physical Review
  A}\ }\textbf {\bibinfo {volume} {67}},\ \bibinfo {pages} {052109} (\bibinfo
  {year} {2003})}\BibitemShut {NoStop}%
\bibitem [{\citenamefont {Deffner}\ and\ \citenamefont
  {Lutz}(2013)}]{deffner2013quantum}%
  \BibitemOpen
  \bibfield  {author} {\bibinfo {author} {\bibfnamefont {S.}~\bibnamefont
  {Deffner}}\ and\ \bibinfo {author} {\bibfnamefont {E.}~\bibnamefont {Lutz}},\
  }\href@noop {} {\bibfield  {journal} {\bibinfo  {journal} {Physical review
  letters}\ }\textbf {\bibinfo {volume} {111}},\ \bibinfo {pages} {010402}
  (\bibinfo {year} {2013})}\BibitemShut {NoStop}%
\bibitem [{\citenamefont {Taddei}\ \emph {et~al.}(2013)\citenamefont {Taddei},
  \citenamefont {Escher}, \citenamefont {Davidovich},\ and\ \citenamefont
  {de~Matos~Filho}}]{taddei2013quantum}%
  \BibitemOpen
  \bibfield  {author} {\bibinfo {author} {\bibfnamefont {M.~M.}\ \bibnamefont
  {Taddei}}, \bibinfo {author} {\bibfnamefont {B.~M.}\ \bibnamefont {Escher}},
  \bibinfo {author} {\bibfnamefont {L.}~\bibnamefont {Davidovich}},\ and\
  \bibinfo {author} {\bibfnamefont {R.~L.}\ \bibnamefont {de~Matos~Filho}},\
  }\href@noop {} {\bibfield  {journal} {\bibinfo  {journal} {Physical review
  letters}\ }\textbf {\bibinfo {volume} {110}},\ \bibinfo {pages} {050402}
  (\bibinfo {year} {2013})}\BibitemShut {NoStop}%
\bibitem [{\citenamefont {del Campo}\ \emph {et~al.}(2013)\citenamefont {del
  Campo}, \citenamefont {Egusquiza}, \citenamefont {Plenio},\ and\
  \citenamefont {Huelga}}]{del2013quantum}%
  \BibitemOpen
  \bibfield  {author} {\bibinfo {author} {\bibfnamefont {A.}~\bibnamefont {del
  Campo}}, \bibinfo {author} {\bibfnamefont {I.~L.}\ \bibnamefont {Egusquiza}},
  \bibinfo {author} {\bibfnamefont {M.~B.}\ \bibnamefont {Plenio}},\ and\
  \bibinfo {author} {\bibfnamefont {S.~F.}\ \bibnamefont {Huelga}},\
  }\href@noop {} {\bibfield  {journal} {\bibinfo  {journal} {Physical review
  letters}\ }\textbf {\bibinfo {volume} {110}},\ \bibinfo {pages} {050403}
  (\bibinfo {year} {2013})}\BibitemShut {NoStop}%
\bibitem [{\citenamefont {Epstein}\ and\ \citenamefont
  {Whaley}(2017)}]{epstein2017quantum}%
  \BibitemOpen
  \bibfield  {author} {\bibinfo {author} {\bibfnamefont {J.~M.}\ \bibnamefont
  {Epstein}}\ and\ \bibinfo {author} {\bibfnamefont {K.~B.}\ \bibnamefont
  {Whaley}},\ }\href@noop {} {\bibfield  {journal} {\bibinfo  {journal}
  {Physical Review A}\ }\textbf {\bibinfo {volume} {95}},\ \bibinfo {pages}
  {042314} (\bibinfo {year} {2017})}\BibitemShut {NoStop}%
\bibitem [{\citenamefont {Deffner}(2020)}]{deffner2020quantum}%
  \BibitemOpen
  \bibfield  {author} {\bibinfo {author} {\bibfnamefont {S.}~\bibnamefont
  {Deffner}},\ }\href@noop {} {\bibfield  {journal} {\bibinfo  {journal}
  {Physical Review Research}\ }\textbf {\bibinfo {volume} {2}},\ \bibinfo
  {pages} {013161} (\bibinfo {year} {2020})}\BibitemShut {NoStop}%
\bibitem [{\citenamefont {Mohan}\ \emph {et~al.}(2022)\citenamefont {Mohan},
  \citenamefont {Das},\ and\ \citenamefont {Pati}}]{mohan2022quantum}%
  \BibitemOpen
  \bibfield  {author} {\bibinfo {author} {\bibfnamefont {B.}~\bibnamefont
  {Mohan}}, \bibinfo {author} {\bibfnamefont {S.}~\bibnamefont {Das}},\ and\
  \bibinfo {author} {\bibfnamefont {A.~K.}\ \bibnamefont {Pati}},\ }\href@noop
  {} {\bibfield  {journal} {\bibinfo  {journal} {New Journal of Physics}\
  }\textbf {\bibinfo {volume} {24}},\ \bibinfo {pages} {065003} (\bibinfo
  {year} {2022})}\BibitemShut {NoStop}%
\bibitem [{\citenamefont {Campaioli}\ \emph {et~al.}(2017)\citenamefont
  {Campaioli}, \citenamefont {Pollock}, \citenamefont {Binder}, \citenamefont
  {C{\'e}leri}, \citenamefont {Goold}, \citenamefont {Vinjanampathy},\ and\
  \citenamefont {Modi}}]{campaioli2017enhancing}%
  \BibitemOpen
  \bibfield  {author} {\bibinfo {author} {\bibfnamefont {F.}~\bibnamefont
  {Campaioli}}, \bibinfo {author} {\bibfnamefont {F.~A.}\ \bibnamefont
  {Pollock}}, \bibinfo {author} {\bibfnamefont {F.~C.}\ \bibnamefont {Binder}},
  \bibinfo {author} {\bibfnamefont {L.}~\bibnamefont {C{\'e}leri}}, \bibinfo
  {author} {\bibfnamefont {J.}~\bibnamefont {Goold}}, \bibinfo {author}
  {\bibfnamefont {S.}~\bibnamefont {Vinjanampathy}},\ and\ \bibinfo {author}
  {\bibfnamefont {K.}~\bibnamefont {Modi}},\ }\href@noop {} {\bibfield
  {journal} {\bibinfo  {journal} {Physical review letters}\ }\textbf {\bibinfo
  {volume} {118}},\ \bibinfo {pages} {150601} (\bibinfo {year}
  {2017})}\BibitemShut {NoStop}%
\bibitem [{\citenamefont {Funo}\ \emph {et~al.}(2019)\citenamefont {Funo},
  \citenamefont {Shiraishi},\ and\ \citenamefont {Saito}}]{funo2019speed}%
  \BibitemOpen
  \bibfield  {author} {\bibinfo {author} {\bibfnamefont {K.}~\bibnamefont
  {Funo}}, \bibinfo {author} {\bibfnamefont {N.}~\bibnamefont {Shiraishi}},\
  and\ \bibinfo {author} {\bibfnamefont {K.}~\bibnamefont {Saito}},\
  }\href@noop {} {\bibfield  {journal} {\bibinfo  {journal} {New Journal of
  Physics}\ }\textbf {\bibinfo {volume} {21}},\ \bibinfo {pages} {013006}
  (\bibinfo {year} {2019})}\BibitemShut {NoStop}%
\bibitem [{\citenamefont {Maleki}\ \emph {et~al.}(2023)\citenamefont {Maleki},
  \citenamefont {Ahansaz},\ and\ \citenamefont {Maleki}}]{maleki2023speed}%
  \BibitemOpen
  \bibfield  {author} {\bibinfo {author} {\bibfnamefont {Y.}~\bibnamefont
  {Maleki}}, \bibinfo {author} {\bibfnamefont {B.}~\bibnamefont {Ahansaz}},\
  and\ \bibinfo {author} {\bibfnamefont {A.}~\bibnamefont {Maleki}},\
  }\href@noop {} {\bibfield  {journal} {\bibinfo  {journal} {Scientific
  Reports}\ }\textbf {\bibinfo {volume} {13}},\ \bibinfo {pages} {12031}
  (\bibinfo {year} {2023})}\BibitemShut {NoStop}%
\bibitem [{\citenamefont {Herb}\ and\ \citenamefont
  {Degen}(2024)}]{herb2024quantum}%
  \BibitemOpen
  \bibfield  {author} {\bibinfo {author} {\bibfnamefont {K.}~\bibnamefont
  {Herb}}\ and\ \bibinfo {author} {\bibfnamefont {C.~L.}\ \bibnamefont
  {Degen}},\ }\href@noop {} {\bibfield  {journal} {\bibinfo  {journal}
  {Physical Review Letters}\ }\textbf {\bibinfo {volume} {133}},\ \bibinfo
  {pages} {210802} (\bibinfo {year} {2024})}\BibitemShut {NoStop}%
\bibitem [{\citenamefont {Breuer}\ and\ \citenamefont
  {Petruccione}(2002)}]{breuer2002theory}%
  \BibitemOpen
  \bibfield  {author} {\bibinfo {author} {\bibfnamefont {H.-P.}\ \bibnamefont
  {Breuer}}\ and\ \bibinfo {author} {\bibfnamefont {F.}~\bibnamefont
  {Petruccione}},\ }\href@noop {} {\emph {\bibinfo {title} {The theory of open
  quantum systems}}}\ (\bibinfo  {publisher} {Oxford University Press, USA},\
  \bibinfo {year} {2002})\BibitemShut {NoStop}%
\bibitem [{\citenamefont {Chanda}\ and\ \citenamefont
  {Bhattacharyya}(2020)}]{chanda2020optimal}%
  \BibitemOpen
  \bibfield  {author} {\bibinfo {author} {\bibfnamefont {N.}~\bibnamefont
  {Chanda}}\ and\ \bibinfo {author} {\bibfnamefont {R.}~\bibnamefont
  {Bhattacharyya}},\ }\href@noop {} {\bibfield  {journal} {\bibinfo  {journal}
  {Physical Review A}\ }\textbf {\bibinfo {volume} {101}},\ \bibinfo {pages}
  {042326} (\bibinfo {year} {2020})}\BibitemShut {NoStop}%
\bibitem [{\citenamefont {Chanda}\ \emph {et~al.}(2023)\citenamefont {Chanda},
  \citenamefont {Patnaik},\ and\ \citenamefont
  {Bhattacharyya}}]{chanda2023optimal}%
  \BibitemOpen
  \bibfield  {author} {\bibinfo {author} {\bibfnamefont {N.}~\bibnamefont
  {Chanda}}, \bibinfo {author} {\bibfnamefont {P.}~\bibnamefont {Patnaik}},\
  and\ \bibinfo {author} {\bibfnamefont {R.}~\bibnamefont {Bhattacharyya}},\
  }\href@noop {} {\bibfield  {journal} {\bibinfo  {journal} {Physical Review
  A}\ }\textbf {\bibinfo {volume} {107}},\ \bibinfo {pages} {063708} (\bibinfo
  {year} {2023})}\BibitemShut {NoStop}%
\bibitem [{\citenamefont {Chakrabarti}\ and\ \citenamefont
  {Bhattacharyya}(2018{\natexlab{a}})}]{frqme}%
  \BibitemOpen
  \bibfield  {author} {\bibinfo {author} {\bibfnamefont {A.}~\bibnamefont
  {Chakrabarti}}\ and\ \bibinfo {author} {\bibfnamefont {R.}~\bibnamefont
  {Bhattacharyya}},\ }\href@noop {} {\bibfield  {journal} {\bibinfo  {journal}
  {Phys. Rev. A}\ }\textbf {\bibinfo {volume} {97}},\ \bibinfo {pages} {063837}
  (\bibinfo {year} {2018}{\natexlab{a}})}\BibitemShut {NoStop}%
\bibitem [{\citenamefont {Wangsness}\ and\ \citenamefont
  {Bloch}(1953)}]{bloch53}%
  \BibitemOpen
  \bibfield  {author} {\bibinfo {author} {\bibfnamefont {R.~K.}\ \bibnamefont
  {Wangsness}}\ and\ \bibinfo {author} {\bibfnamefont {F.}~\bibnamefont
  {Bloch}},\ }\href@noop {} {\bibfield  {journal} {\bibinfo  {journal} {Phys.
  Rev.}\ }\textbf {\bibinfo {volume} {89}},\ \bibinfo {pages} {728} (\bibinfo
  {year} {1953})}\BibitemShut {NoStop}%
\bibitem [{\citenamefont {Redfield}(1957)}]{redfield}%
  \BibitemOpen
  \bibfield  {author} {\bibinfo {author} {\bibfnamefont {A.~G.}\ \bibnamefont
  {Redfield}},\ }\href@noop {} {\bibfield  {journal} {\bibinfo  {journal} {IBM
  Journal of Research and Development}\ }\textbf {\bibinfo {volume} {1}},\
  \bibinfo {pages} {19} (\bibinfo {year} {1957})}\BibitemShut {NoStop}%
\bibitem [{\citenamefont {Gorini}\ \emph {et~al.}(1976)\citenamefont {Gorini},
  \citenamefont {Kossakowski},\ and\ \citenamefont {Sudarshan}}]{lindblad}%
  \BibitemOpen
  \bibfield  {author} {\bibinfo {author} {\bibfnamefont {V.}~\bibnamefont
  {Gorini}}, \bibinfo {author} {\bibfnamefont {A.}~\bibnamefont
  {Kossakowski}},\ and\ \bibinfo {author} {\bibfnamefont {E.~C.~G.}\
  \bibnamefont {Sudarshan}},\ }\href@noop {} {\bibfield  {journal} {\bibinfo
  {journal} {Journal of Mathematical Physics}\ }\textbf {\bibinfo {volume}
  {17}},\ \bibinfo {pages} {821} (\bibinfo {year} {1976})}\BibitemShut
  {NoStop}%
\bibitem [{\citenamefont {Saha}\ and\ \citenamefont
  {Bhattacharyya}(2024{\natexlab{a}})}]{saha2024applications}%
  \BibitemOpen
  \bibfield  {author} {\bibinfo {author} {\bibfnamefont {S.}~\bibnamefont
  {Saha}}\ and\ \bibinfo {author} {\bibfnamefont {R.}~\bibnamefont
  {Bhattacharyya}},\ }\href@noop {} {\bibfield  {journal} {\bibinfo  {journal}
  {The European Physical Journal Special Topics}\ ,\ \bibinfo {pages} {1}}
  (\bibinfo {year} {2024}{\natexlab{a}})}\BibitemShut {NoStop}%
\bibitem [{\citenamefont {Chatterjee}\ and\ \citenamefont
  {Bhattacharyya}(2024)}]{chatterjee2024improved}%
  \BibitemOpen
  \bibfield  {author} {\bibinfo {author} {\bibfnamefont {A.}~\bibnamefont
  {Chatterjee}}\ and\ \bibinfo {author} {\bibfnamefont {R.}~\bibnamefont
  {Bhattacharyya}},\ }\href@noop {} {\bibfield  {journal} {\bibinfo  {journal}
  {The European Physical Journal D}\ }\textbf {\bibinfo {volume} {78}},\
  \bibinfo {pages} {44} (\bibinfo {year} {2024})}\BibitemShut {NoStop}%
\bibitem [{\citenamefont {Chanda}\ and\ \citenamefont
  {Bhattacharyya}(2021)}]{chanda2021emergence}%
  \BibitemOpen
  \bibfield  {author} {\bibinfo {author} {\bibfnamefont {N.}~\bibnamefont
  {Chanda}}\ and\ \bibinfo {author} {\bibfnamefont {R.}~\bibnamefont
  {Bhattacharyya}},\ }\href@noop {} {\bibfield  {journal} {\bibinfo  {journal}
  {Physical Review A}\ }\textbf {\bibinfo {volume} {104}},\ \bibinfo {pages}
  {022436} (\bibinfo {year} {2021})}\BibitemShut {NoStop}%
\bibitem [{\citenamefont {Das}\ and\ \citenamefont
  {Bhattacharyya}(2024)}]{das2024irreversibility}%
  \BibitemOpen
  \bibfield  {author} {\bibinfo {author} {\bibfnamefont {G.}~\bibnamefont
  {Das}}\ and\ \bibinfo {author} {\bibfnamefont {R.}~\bibnamefont
  {Bhattacharyya}},\ }\href@noop {} {\bibfield  {journal} {\bibinfo  {journal}
  {Physical Review A}\ }\textbf {\bibinfo {volume} {110}},\ \bibinfo {pages}
  {062211} (\bibinfo {year} {2024})}\BibitemShut {NoStop}%
\bibitem [{\citenamefont {Saha}\ and\ \citenamefont
  {Bhattacharyya}(2023)}]{saha2023cascaded}%
  \BibitemOpen
  \bibfield  {author} {\bibinfo {author} {\bibfnamefont {S.}~\bibnamefont
  {Saha}}\ and\ \bibinfo {author} {\bibfnamefont {R.}~\bibnamefont
  {Bhattacharyya}},\ }\href@noop {} {\bibfield  {journal} {\bibinfo  {journal}
  {Physical Review A}\ }\textbf {\bibinfo {volume} {107}},\ \bibinfo {pages}
  {022206} (\bibinfo {year} {2023})}\BibitemShut {NoStop}%
\bibitem [{\citenamefont {Saha}\ and\ \citenamefont
  {Bhattacharyya}(2024{\natexlab{b}})}]{saha2024prethermalization}%
  \BibitemOpen
  \bibfield  {author} {\bibinfo {author} {\bibfnamefont {S.}~\bibnamefont
  {Saha}}\ and\ \bibinfo {author} {\bibfnamefont {R.}~\bibnamefont
  {Bhattacharyya}},\ }\href@noop {} {\bibfield  {journal} {\bibinfo  {journal}
  {Journal of Statistical Mechanics: Theory and Experiment}\ }\textbf {\bibinfo
  {volume} {2024}},\ \bibinfo {pages} {023103} (\bibinfo {year}
  {2024}{\natexlab{b}})}\BibitemShut {NoStop}%
\bibitem [{\citenamefont {Saha}\ and\ \citenamefont
  {Bhattacharyya}(2024{\natexlab{c}})}]{saha2024prethermal}%
  \BibitemOpen
  \bibfield  {author} {\bibinfo {author} {\bibfnamefont {S.}~\bibnamefont
  {Saha}}\ and\ \bibinfo {author} {\bibfnamefont {R.}~\bibnamefont
  {Bhattacharyya}},\ }\href@noop {} {\bibfield  {journal} {\bibinfo  {journal}
  {Physical Review A}\ }\textbf {\bibinfo {volume} {109}},\ \bibinfo {pages}
  {012208} (\bibinfo {year} {2024}{\natexlab{c}})}\BibitemShut {NoStop}%
\bibitem [{\citenamefont {Das}\ \emph {et~al.}(2025)\citenamefont {Das},
  \citenamefont {Saha},\ and\ \citenamefont {Bhattacharyya}}]{das2025discrete}%
  \BibitemOpen
  \bibfield  {author} {\bibinfo {author} {\bibfnamefont {G.}~\bibnamefont
  {Das}}, \bibinfo {author} {\bibfnamefont {S.}~\bibnamefont {Saha}},\ and\
  \bibinfo {author} {\bibfnamefont {R.}~\bibnamefont {Bhattacharyya}},\
  }\href@noop {} {\bibfield  {journal} {\bibinfo  {journal} {arXiv preprint
  arXiv:2511.09852}\ } (\bibinfo {year} {2025})}\BibitemShut {NoStop}%
\bibitem [{\citenamefont {Nielsen}\ \emph {et~al.}(2006)\citenamefont
  {Nielsen}, \citenamefont {Dowling}, \citenamefont {Gu},\ and\ \citenamefont
  {Doherty}}]{nielsen2006optimal}%
  \BibitemOpen
  \bibfield  {author} {\bibinfo {author} {\bibfnamefont {M.~A.}\ \bibnamefont
  {Nielsen}}, \bibinfo {author} {\bibfnamefont {M.~R.}\ \bibnamefont
  {Dowling}}, \bibinfo {author} {\bibfnamefont {M.}~\bibnamefont {Gu}},\ and\
  \bibinfo {author} {\bibfnamefont {A.~C.}\ \bibnamefont {Doherty}},\
  }\href@noop {} {\bibfield  {journal} {\bibinfo  {journal} {Physical Review
  A—Atomic, Molecular, and Optical Physics}\ }\textbf {\bibinfo {volume}
  {73}},\ \bibinfo {pages} {062323} (\bibinfo {year} {2006})}\BibitemShut
  {NoStop}%
\bibitem [{\citenamefont {Werschnik}\ and\ \citenamefont
  {Gross}(2007)}]{werschnik2007quantum}%
  \BibitemOpen
  \bibfield  {author} {\bibinfo {author} {\bibfnamefont {J.}~\bibnamefont
  {Werschnik}}\ and\ \bibinfo {author} {\bibfnamefont {E.}~\bibnamefont
  {Gross}},\ }\href@noop {} {\bibfield  {journal} {\bibinfo  {journal} {Journal
  of Physics B: Atomic, Molecular and Optical Physics}\ }\textbf {\bibinfo
  {volume} {40}},\ \bibinfo {pages} {R175} (\bibinfo {year}
  {2007})}\BibitemShut {NoStop}%
\bibitem [{\citenamefont {Khaneja}\ \emph {et~al.}(2005)\citenamefont
  {Khaneja}, \citenamefont {Reiss}, \citenamefont {Kehlet}, \citenamefont
  {Schulte-Herbr{\"u}ggen},\ and\ \citenamefont {Glaser}}]{khaneja2005optimal}%
  \BibitemOpen
  \bibfield  {author} {\bibinfo {author} {\bibfnamefont {N.}~\bibnamefont
  {Khaneja}}, \bibinfo {author} {\bibfnamefont {T.}~\bibnamefont {Reiss}},
  \bibinfo {author} {\bibfnamefont {C.}~\bibnamefont {Kehlet}}, \bibinfo
  {author} {\bibfnamefont {T.}~\bibnamefont {Schulte-Herbr{\"u}ggen}},\ and\
  \bibinfo {author} {\bibfnamefont {S.~J.}\ \bibnamefont {Glaser}},\
  }\href@noop {} {\bibfield  {journal} {\bibinfo  {journal} {Journal of
  magnetic resonance}\ }\textbf {\bibinfo {volume} {172}},\ \bibinfo {pages}
  {296} (\bibinfo {year} {2005})}\BibitemShut {NoStop}%
\bibitem [{\citenamefont {Schulte-Herbr{\"u}ggen}\ \emph
  {et~al.}(2011)\citenamefont {Schulte-Herbr{\"u}ggen}, \citenamefont
  {Sp{\"o}rl}, \citenamefont {Khaneja},\ and\ \citenamefont
  {Glaser}}]{schulte2011optimal}%
  \BibitemOpen
  \bibfield  {author} {\bibinfo {author} {\bibfnamefont {T.}~\bibnamefont
  {Schulte-Herbr{\"u}ggen}}, \bibinfo {author} {\bibfnamefont {A.}~\bibnamefont
  {Sp{\"o}rl}}, \bibinfo {author} {\bibfnamefont {N.}~\bibnamefont {Khaneja}},\
  and\ \bibinfo {author} {\bibfnamefont {S.}~\bibnamefont {Glaser}},\
  }\href@noop {} {\bibfield  {journal} {\bibinfo  {journal} {Journal of Physics
  B: Atomic, Molecular and Optical Physics}\ }\textbf {\bibinfo {volume}
  {44}},\ \bibinfo {pages} {154013} (\bibinfo {year} {2011})}\BibitemShut
  {NoStop}%
\bibitem [{\citenamefont {Caneva}\ \emph {et~al.}(2009)\citenamefont {Caneva},
  \citenamefont {Murphy}, \citenamefont {Calarco}, \citenamefont {Fazio},
  \citenamefont {Montangero}, \citenamefont {Giovannetti},\ and\ \citenamefont
  {Santoro}}]{caneva2009optimal}%
  \BibitemOpen
  \bibfield  {author} {\bibinfo {author} {\bibfnamefont {T.}~\bibnamefont
  {Caneva}}, \bibinfo {author} {\bibfnamefont {M.}~\bibnamefont {Murphy}},
  \bibinfo {author} {\bibfnamefont {T.}~\bibnamefont {Calarco}}, \bibinfo
  {author} {\bibfnamefont {R.}~\bibnamefont {Fazio}}, \bibinfo {author}
  {\bibfnamefont {S.}~\bibnamefont {Montangero}}, \bibinfo {author}
  {\bibfnamefont {V.}~\bibnamefont {Giovannetti}},\ and\ \bibinfo {author}
  {\bibfnamefont {G.~E.}\ \bibnamefont {Santoro}},\ }\href@noop {} {\bibfield
  {journal} {\bibinfo  {journal} {Physical review letters}\ }\textbf {\bibinfo
  {volume} {103}},\ \bibinfo {pages} {240501} (\bibinfo {year}
  {2009})}\BibitemShut {NoStop}%
\bibitem [{\citenamefont {Poggi}\ \emph {et~al.}(2013)\citenamefont {Poggi},
  \citenamefont {Lombardo},\ and\ \citenamefont
  {Wisniacki}}]{poggi2013quantum}%
  \BibitemOpen
  \bibfield  {author} {\bibinfo {author} {\bibfnamefont {P.~M.}\ \bibnamefont
  {Poggi}}, \bibinfo {author} {\bibfnamefont {F.~C.}\ \bibnamefont
  {Lombardo}},\ and\ \bibinfo {author} {\bibfnamefont {D.}~\bibnamefont
  {Wisniacki}},\ }\href@noop {} {\bibfield  {journal} {\bibinfo  {journal}
  {Europhysics Letters}\ }\textbf {\bibinfo {volume} {104}},\ \bibinfo {pages}
  {40005} (\bibinfo {year} {2013})}\BibitemShut {NoStop}%
\bibitem [{\citenamefont {Brouzos}\ \emph {et~al.}(2015)\citenamefont
  {Brouzos}, \citenamefont {Streltsov}, \citenamefont {Negretti}, \citenamefont
  {Said}, \citenamefont {Caneva}, \citenamefont {Montangero},\ and\
  \citenamefont {Calarco}}]{brouzos2015quantum}%
  \BibitemOpen
  \bibfield  {author} {\bibinfo {author} {\bibfnamefont {I.}~\bibnamefont
  {Brouzos}}, \bibinfo {author} {\bibfnamefont {A.~I.}\ \bibnamefont
  {Streltsov}}, \bibinfo {author} {\bibfnamefont {A.}~\bibnamefont {Negretti}},
  \bibinfo {author} {\bibfnamefont {R.~S.}\ \bibnamefont {Said}}, \bibinfo
  {author} {\bibfnamefont {T.}~\bibnamefont {Caneva}}, \bibinfo {author}
  {\bibfnamefont {S.}~\bibnamefont {Montangero}},\ and\ \bibinfo {author}
  {\bibfnamefont {T.}~\bibnamefont {Calarco}},\ }\href@noop {} {\bibfield
  {journal} {\bibinfo  {journal} {Physical Review A}\ }\textbf {\bibinfo
  {volume} {92}},\ \bibinfo {pages} {062110} (\bibinfo {year}
  {2015})}\BibitemShut {NoStop}%
\bibitem [{\citenamefont {Lucarelli}(2018)}]{PhysRevA.97.062346}%
  \BibitemOpen
  \bibfield  {author} {\bibinfo {author} {\bibfnamefont {D.}~\bibnamefont
  {Lucarelli}},\ }\href@noop {} {\bibfield  {journal} {\bibinfo  {journal}
  {Phys. Rev. A}\ }\textbf {\bibinfo {volume} {97}},\ \bibinfo {pages} {062346}
  (\bibinfo {year} {2018})}\BibitemShut {NoStop}%
\bibitem [{\citenamefont {Cao}\ \emph {et~al.}(2024)\citenamefont {Cao},
  \citenamefont {Cui}, \citenamefont {Yung},\ and\ \citenamefont
  {Wu}}]{cao2024robust}%
  \BibitemOpen
  \bibfield  {author} {\bibinfo {author} {\bibfnamefont {X.}~\bibnamefont
  {Cao}}, \bibinfo {author} {\bibfnamefont {J.}~\bibnamefont {Cui}}, \bibinfo
  {author} {\bibfnamefont {M.~H.}\ \bibnamefont {Yung}},\ and\ \bibinfo
  {author} {\bibfnamefont {R.-B.}\ \bibnamefont {Wu}},\ }\href@noop {}
  {\bibfield  {journal} {\bibinfo  {journal} {Physical Review A}\ }\textbf
  {\bibinfo {volume} {110}},\ \bibinfo {pages} {022603} (\bibinfo {year}
  {2024})}\BibitemShut {NoStop}%
\bibitem [{\citenamefont {Impens}\ and\ \citenamefont
  {Gu{\'e}ry-Odelin}(2025)}]{impens2025approaching}%
  \BibitemOpen
  \bibfield  {author} {\bibinfo {author} {\bibfnamefont {F.}~\bibnamefont
  {Impens}}\ and\ \bibinfo {author} {\bibfnamefont {D.}~\bibnamefont
  {Gu{\'e}ry-Odelin}},\ }\href@noop {} {\bibfield  {journal} {\bibinfo
  {journal} {Physical Review A}\ }\textbf {\bibinfo {volume} {112}},\ \bibinfo
  {pages} {042614} (\bibinfo {year} {2025})}\BibitemShut {NoStop}%
\bibitem [{\citenamefont {Haeberlen}(2012)}]{secular_approx}%
  \BibitemOpen
  \bibfield  {author} {\bibinfo {author} {\bibfnamefont {U.}~\bibnamefont
  {Haeberlen}},\ }\href@noop {} {\emph {\bibinfo {title} {High Resolution NMR
  in solids selective averaging: supplement 1 advances in magnetic
  resonance}}},\ Vol.~\bibinfo {volume} {1}\ (\bibinfo  {publisher}
  {Elsevier},\ \bibinfo {year} {2012})\BibitemShut {NoStop}%
\bibitem [{\citenamefont {Chakrabarti}\ and\ \citenamefont
  {Bhattacharyya}(2018{\natexlab{b}})}]{didexpt}%
  \BibitemOpen
  \bibfield  {author} {\bibinfo {author} {\bibfnamefont {A.}~\bibnamefont
  {Chakrabarti}}\ and\ \bibinfo {author} {\bibfnamefont {R.}~\bibnamefont
  {Bhattacharyya}},\ }\href@noop {} {\bibfield  {journal} {\bibinfo  {journal}
  {Europhysics Letters}\ }\textbf {\bibinfo {volume} {121}},\ \bibinfo {pages}
  {57002} (\bibinfo {year} {2018}{\natexlab{b}})}\BibitemShut {NoStop}%
\bibitem [{\citenamefont {Konnov}\ and\ \citenamefont
  {Krotov}(1999)}]{konnov1999global}%
  \BibitemOpen
  \bibfield  {author} {\bibinfo {author} {\bibfnamefont {A.~I.}\ \bibnamefont
  {Konnov}}\ and\ \bibinfo {author} {\bibfnamefont {V.~F.}\ \bibnamefont
  {Krotov}},\ }\href@noop {} {\bibfield  {journal} {\bibinfo  {journal}
  {Avtomatika i Telemekhanika}\ ,\ \bibinfo {pages} {77}} (\bibinfo {year}
  {1999})}\BibitemShut {NoStop}%
\bibitem [{\citenamefont {Caneva}\ \emph {et~al.}(2011)\citenamefont {Caneva},
  \citenamefont {Calarco},\ and\ \citenamefont
  {Montangero}}]{caneva2011chopped}%
  \BibitemOpen
  \bibfield  {author} {\bibinfo {author} {\bibfnamefont {T.}~\bibnamefont
  {Caneva}}, \bibinfo {author} {\bibfnamefont {T.}~\bibnamefont {Calarco}},\
  and\ \bibinfo {author} {\bibfnamefont {S.}~\bibnamefont {Montangero}},\
  }\href@noop {} {\bibfield  {journal} {\bibinfo  {journal} {Physical Review
  A—Atomic, Molecular, and Optical Physics}\ }\textbf {\bibinfo {volume}
  {84}},\ \bibinfo {pages} {022326} (\bibinfo {year} {2011})}\BibitemShut
  {NoStop}%
\bibitem [{\citenamefont {Machnes}\ \emph {et~al.}(2011)\citenamefont
  {Machnes}, \citenamefont {Sander}, \citenamefont {Glaser}, \citenamefont
  {de~Fouquieres}, \citenamefont {Gruslys}, \citenamefont {Schirmer},\ and\
  \citenamefont {Schulte-Herbr{\"u}ggen}}]{machnes2011comparing}%
  \BibitemOpen
  \bibfield  {author} {\bibinfo {author} {\bibfnamefont {S.}~\bibnamefont
  {Machnes}}, \bibinfo {author} {\bibfnamefont {U.}~\bibnamefont {Sander}},
  \bibinfo {author} {\bibfnamefont {S.~J.}\ \bibnamefont {Glaser}}, \bibinfo
  {author} {\bibfnamefont {P.}~\bibnamefont {de~Fouquieres}}, \bibinfo {author}
  {\bibfnamefont {A.}~\bibnamefont {Gruslys}}, \bibinfo {author} {\bibfnamefont
  {S.}~\bibnamefont {Schirmer}},\ and\ \bibinfo {author} {\bibfnamefont
  {T.}~\bibnamefont {Schulte-Herbr{\"u}ggen}},\ }\href@noop {} {\bibfield
  {journal} {\bibinfo  {journal} {Physical Review A—Atomic, Molecular, and
  Optical Physics}\ }\textbf {\bibinfo {volume} {84}},\ \bibinfo {pages}
  {022305} (\bibinfo {year} {2011})}\BibitemShut {NoStop}%
\bibitem [{\citenamefont {Rabitz}\ \emph {et~al.}(2004)\citenamefont {Rabitz},
  \citenamefont {Hsieh},\ and\ \citenamefont {Rosenthal}}]{rabitz2004quantum}%
  \BibitemOpen
  \bibfield  {author} {\bibinfo {author} {\bibfnamefont {H.~A.}\ \bibnamefont
  {Rabitz}}, \bibinfo {author} {\bibfnamefont {M.~M.}\ \bibnamefont {Hsieh}},\
  and\ \bibinfo {author} {\bibfnamefont {C.~M.}\ \bibnamefont {Rosenthal}},\
  }\href@noop {} {\bibfield  {journal} {\bibinfo  {journal} {Science}\ }\textbf
  {\bibinfo {volume} {303}},\ \bibinfo {pages} {1998} (\bibinfo {year}
  {2004})}\BibitemShut {NoStop}%
\bibitem [{\citenamefont {Bason}\ \emph {et~al.}(2012)\citenamefont {Bason},
  \citenamefont {Viteau}, \citenamefont {Malossi}, \citenamefont {Huillery},
  \citenamefont {Arimondo}, \citenamefont {Ciampini}, \citenamefont {Fazio},
  \citenamefont {Giovannetti}, \citenamefont {Mannella},\ and\ \citenamefont
  {Morsch}}]{bason2012high}%
  \BibitemOpen
  \bibfield  {author} {\bibinfo {author} {\bibfnamefont {M.~G.}\ \bibnamefont
  {Bason}}, \bibinfo {author} {\bibfnamefont {M.}~\bibnamefont {Viteau}},
  \bibinfo {author} {\bibfnamefont {N.}~\bibnamefont {Malossi}}, \bibinfo
  {author} {\bibfnamefont {P.}~\bibnamefont {Huillery}}, \bibinfo {author}
  {\bibfnamefont {E.}~\bibnamefont {Arimondo}}, \bibinfo {author}
  {\bibfnamefont {D.}~\bibnamefont {Ciampini}}, \bibinfo {author}
  {\bibfnamefont {R.}~\bibnamefont {Fazio}}, \bibinfo {author} {\bibfnamefont
  {V.}~\bibnamefont {Giovannetti}}, \bibinfo {author} {\bibfnamefont
  {R.}~\bibnamefont {Mannella}},\ and\ \bibinfo {author} {\bibfnamefont
  {O.}~\bibnamefont {Morsch}},\ }\href@noop {} {\bibfield  {journal} {\bibinfo
  {journal} {Nature Physics}\ }\textbf {\bibinfo {volume} {8}},\ \bibinfo
  {pages} {147} (\bibinfo {year} {2012})}\BibitemShut {NoStop}%
\bibitem [{\citenamefont {Wang}\ \emph {et~al.}(2017)\citenamefont {Wang},
  \citenamefont {Um}, \citenamefont {Zhang}, \citenamefont {An}, \citenamefont
  {Lyu}, \citenamefont {Zhang}, \citenamefont {Duan}, \citenamefont {Yum},\
  and\ \citenamefont {Kim}}]{wang2017single}%
  \BibitemOpen
  \bibfield  {author} {\bibinfo {author} {\bibfnamefont {Y.}~\bibnamefont
  {Wang}}, \bibinfo {author} {\bibfnamefont {M.}~\bibnamefont {Um}}, \bibinfo
  {author} {\bibfnamefont {J.}~\bibnamefont {Zhang}}, \bibinfo {author}
  {\bibfnamefont {S.}~\bibnamefont {An}}, \bibinfo {author} {\bibfnamefont
  {M.}~\bibnamefont {Lyu}}, \bibinfo {author} {\bibfnamefont {J.-N.}\
  \bibnamefont {Zhang}}, \bibinfo {author} {\bibfnamefont {L.-M.}\ \bibnamefont
  {Duan}}, \bibinfo {author} {\bibfnamefont {D.}~\bibnamefont {Yum}},\ and\
  \bibinfo {author} {\bibfnamefont {K.}~\bibnamefont {Kim}},\ }\href@noop {}
  {\bibfield  {journal} {\bibinfo  {journal} {Nature Photonics}\ }\textbf
  {\bibinfo {volume} {11}},\ \bibinfo {pages} {646} (\bibinfo {year}
  {2017})}\BibitemShut {NoStop}%
\bibitem [{\citenamefont {Devoret}\ and\ \citenamefont
  {Schoelkopf}(2013)}]{devoret2013superconducting}%
  \BibitemOpen
  \bibfield  {author} {\bibinfo {author} {\bibfnamefont {M.~H.}\ \bibnamefont
  {Devoret}}\ and\ \bibinfo {author} {\bibfnamefont {R.~J.}\ \bibnamefont
  {Schoelkopf}},\ }\href@noop {} {\bibfield  {journal} {\bibinfo  {journal}
  {Science}\ }\textbf {\bibinfo {volume} {339}},\ \bibinfo {pages} {1169}
  (\bibinfo {year} {2013})}\BibitemShut {NoStop}%
\bibitem [{\citenamefont {Ebert}\ \emph {et~al.}(2015)\citenamefont {Ebert},
  \citenamefont {Kwon}, \citenamefont {Walker},\ and\ \citenamefont
  {Saffman}}]{ebert2015coherence}%
  \BibitemOpen
  \bibfield  {author} {\bibinfo {author} {\bibfnamefont {M.}~\bibnamefont
  {Ebert}}, \bibinfo {author} {\bibfnamefont {M.}~\bibnamefont {Kwon}},
  \bibinfo {author} {\bibfnamefont {T.}~\bibnamefont {Walker}},\ and\ \bibinfo
  {author} {\bibfnamefont {M.}~\bibnamefont {Saffman}},\ }\href@noop {}
  {\bibfield  {journal} {\bibinfo  {journal} {Physical Review Letters}\
  }\textbf {\bibinfo {volume} {115}},\ \bibinfo {pages} {093601} (\bibinfo
  {year} {2015})}\BibitemShut {NoStop}%
\bibitem [{\citenamefont {Veldhorst}\ \emph {et~al.}(2014)\citenamefont
  {Veldhorst}, \citenamefont {Hwang}, \citenamefont {Yang}, \citenamefont
  {Leenstra}, \citenamefont {de~Ronde}, \citenamefont {Dehollain},
  \citenamefont {Muhonen}, \citenamefont {Hudson}, \citenamefont {Itoh},
  \citenamefont {Morello} \emph {et~al.}}]{veldhorst2014addressable}%
  \BibitemOpen
  \bibfield  {author} {\bibinfo {author} {\bibfnamefont {M.}~\bibnamefont
  {Veldhorst}}, \bibinfo {author} {\bibfnamefont {J.~C.}\ \bibnamefont
  {Hwang}}, \bibinfo {author} {\bibfnamefont {C.-H.}\ \bibnamefont {Yang}},
  \bibinfo {author} {\bibfnamefont {A.~W.}\ \bibnamefont {Leenstra}}, \bibinfo
  {author} {\bibfnamefont {B.}~\bibnamefont {de~Ronde}}, \bibinfo {author}
  {\bibfnamefont {J.~P.}\ \bibnamefont {Dehollain}}, \bibinfo {author}
  {\bibfnamefont {J.~T.}\ \bibnamefont {Muhonen}}, \bibinfo {author}
  {\bibfnamefont {F.~E.}\ \bibnamefont {Hudson}}, \bibinfo {author}
  {\bibfnamefont {K.~M.}\ \bibnamefont {Itoh}}, \bibinfo {author}
  {\bibfnamefont {A.}~\bibnamefont {Morello}}, \emph {et~al.},\ }\href@noop {}
  {\bibfield  {journal} {\bibinfo  {journal} {Nature nanotechnology}\ }\textbf
  {\bibinfo {volume} {9}},\ \bibinfo {pages} {981} (\bibinfo {year}
  {2014})}\BibitemShut {NoStop}%
\bibitem [{\citenamefont {Doherty}\ \emph {et~al.}(2013)\citenamefont
  {Doherty}, \citenamefont {Manson}, \citenamefont {Delaney}, \citenamefont
  {Jelezko}, \citenamefont {Wrachtrup},\ and\ \citenamefont
  {Hollenberg}}]{doherty2013nitrogen}%
  \BibitemOpen
  \bibfield  {author} {\bibinfo {author} {\bibfnamefont {M.~W.}\ \bibnamefont
  {Doherty}}, \bibinfo {author} {\bibfnamefont {N.~B.}\ \bibnamefont {Manson}},
  \bibinfo {author} {\bibfnamefont {P.}~\bibnamefont {Delaney}}, \bibinfo
  {author} {\bibfnamefont {F.}~\bibnamefont {Jelezko}}, \bibinfo {author}
  {\bibfnamefont {J.}~\bibnamefont {Wrachtrup}},\ and\ \bibinfo {author}
  {\bibfnamefont {L.~C.}\ \bibnamefont {Hollenberg}},\ }\href@noop {}
  {\bibfield  {journal} {\bibinfo  {journal} {Physics Reports}\ }\textbf
  {\bibinfo {volume} {528}},\ \bibinfo {pages} {1} (\bibinfo {year}
  {2013})}\BibitemShut {NoStop}%
\bibitem [{\citenamefont {Kobayashi}(2025)}]{kobayashi2025quantum}%
  \BibitemOpen
  \bibfield  {author} {\bibinfo {author} {\bibfnamefont {K.}~\bibnamefont
  {Kobayashi}},\ }\href@noop {} {\bibfield  {journal} {\bibinfo  {journal}
  {Journal of Physics Communications}\ }\textbf {\bibinfo {volume} {9}},\
  \bibinfo {pages} {115002} (\bibinfo {year} {2025})}\BibitemShut {NoStop}%
\end{thebibliography}%


\appendix

\section{Analytical bound} \label{Sec: Analytical QSL}

We adapt the bounding technique of Kobayashi \cite{kobayashi2025quantum} to derive an analytical QSL bound using FRQME, extending it to include both environmental Lindblad dissipation and DID. 

To calculate the overlap between two states, $\eta$ and $\rho$, often fidelity is calculated which is given as: 
\begin{equation} \label{Fidelity}
	\text{F}(\eta, \rho) = \Big( \text{Tr}\Big(\sqrt{\sqrt{\eta} \, \rho \, \sqrt{\eta}} \;\;\Big) \Big)^2
\end{equation}

Let's say $\eta$ is pure. This implies Idempotency ($\eta^2 = \eta \implies \sqrt{\eta} = \eta$) which simplifies Eq. \ref{Fidelity} to $$\text{F} = \Big{(} \text{Tr} (\sqrt{\eta \rho \eta} ) \Big{)}^2$$

For a rank $1$ projector $P = \vert \psi \rangle \langle \psi \vert$ ($P^2 = P$; $\text{Tr}(P) = 1$; $P^\dagger = P$) and any matrix $A$: 
\begin{eqnarray}
	P A P =\vert \psi \rangle \langle \psi \vert A \vert \psi \rangle \langle \psi \vert 
	= \langle \psi \vert A \vert \psi \rangle \vert \psi \rangle  \langle \psi \vert 
	= \text{Tr} (P A) P  \nn
\end{eqnarray}

Indeed, $\eta$ is a rank $1$ projector, which can easily be verified. Then, $\eta \rho \eta = \text{Tr}(\eta \rho) \eta \implies X = \lambda \eta$, where, $X = \eta \rho \eta$ and $\lambda = \text{Tr}(\eta \rho)$

Now, 
\begin{eqnarray}
	\sqrt{X} = \sqrt{\lambda \eta} &=& \sqrt{\lambda}\, \eta \nn \\
	\implies \text{Tr} (\sqrt{X}) = \sqrt{\lambda} \, \text{Tr} (\eta) &=& \sqrt{\lambda} \nn \\
	\implies \Big{(}\text{Tr} (\sqrt{X}) \Big{)}^2 &=&  \lambda \nn \\ 
	\implies \text{F} = \Big{(}\text{Tr} (\sqrt{\eta \rho \eta}) \Big{)}^2 &=&  \text{Tr}(\eta \rho)  \nn
\end{eqnarray}

Under the assumption $\eta$ is pure, the fidelity reduces to: 
\begin{eqnarray} \label{F and theta}
	\text{F}(\eta, \rho) = \text{Tr} \big(\eta \rho \big) = \cos^2 \theta
\end{eqnarray}

where $\theta$ is the Bures angle. We take the derivative with respect to time: 
\begin{eqnarray} \label{F dot with theta}
	\dot{F}(\eta, \rho) = -2 \cos \theta \sin \theta \,\dot{\theta} = \text{Tr}(\eta \dot{\rho}) 
\end{eqnarray}

We use the dynamical equation given in Eq. \ref{Dynamical_Equation} with a general system Hamiltonian, take the absolute value to find the upper bound and use the generalized triangle inequality: 
\begin{eqnarray} \label{F dot}
	\vert \dot{F}(\eta, \rho) \vert &\le& \Big{\vert} \text{Tr}\Big{(}- i \, \eta \, [H, \, \rho] \Big{)} \Big{\vert} \nn \\ 
	&+& \Big{\vert} R_1\, \text{Tr}\Big{(} \eta \, \mathcal{D}[\sigma_{+}](\rho)\Big{)} \Big{\vert} + \Big{\vert} R_2\, \text{Tr}\Big{(} \eta \, \mathcal{D}[\sigma_{-}](\rho) \Big{)} \Big{\vert}  \nn  \\ 
	&+& \Big{\vert} \text{Tr}\Big{(} - \int_{0}^{\infty} d \tau \, e^{-\frac{\tau}{\tau_c}} \eta \, [H(t), [H(t- \tau), \rho]] \Big{)} \Big{\vert}
\end{eqnarray}
where, $R_{1,2} = \omega_{\rm SL}^2 \tau_c P_{1, 2}$. The first line of Eq. \ref{F dot} gives the unitary contribution of the system Hamiltonian, the second line gives the environmental dissipation in the Lindblad form for the two-level system, and the third line gives the second-order contributions of the system Hamiltonian. 

Then, using the Cauchy-Schwarz inequality:
\begin{eqnarray} \label{CS inequality}
	\Big{\vert} \text{Tr} (X^\dagger Y) \Big{\vert} \le \Big{\vert} \Big{\vert}  X \Big{\vert} \Big{\vert} _F \, \Big{\vert} \Big{\vert}  Y \Big{\vert} \Big{\vert} _F
\end{eqnarray}
and Frobenius norm: 
\begin{eqnarray} \label{F norm}
	\Big{\vert} \Big{\vert}  X \Big{\vert} \Big{\vert} _F = \sqrt{\text{Tr}(X^\dagger X)}
\end{eqnarray}

The first term can be written as follows: 
\begin{eqnarray} \label{1st term}
	\Big{\vert} \text{Tr}\Big{(} - i \, \eta \, [H, \, \rho] \Big{)} \Big{\vert} &=& \Big{\vert} \,\text{Tr}\Big{(}i[H, \eta]  (\rho - \eta) \Big{)} + \text{Tr}\Big{(} i[H, \eta]  \eta \Big{)} \Big{\vert} \nn \\ 
	&=& \Big{\vert} \,\text{Tr}\Big{(}i[H, \eta]  (\rho - \eta) \Big{)} \Big{\vert}
\end{eqnarray}

The second term is zero. Using Eq. \ref{CS inequality} and Eq. \ref{F norm} on Eq. \ref{1st term}, we get: 
\begin{eqnarray} \label{1st term 1}
	&& \Big{\vert} \,\text{Tr}\Big{(} i [H, \eta]  (\rho - \eta) \Big{)} \Big{\vert} \le \Big{\vert} \Big{\vert}  i [H, \eta]^\dagger \Big{\vert} \Big{\vert} _F \Big{\vert} \Big{\vert}  \rho - \eta \Big{\vert} \Big{\vert} _F \nn \\ 
	&=& \sqrt{\text{Tr}\Big{(} ( i [H, \eta]^\dagger)^\dagger ( i  [H, \eta]^\dagger)\Big{)}} \sqrt{\text{Tr}\Big{(} (\rho - \eta)^\dagger (\rho - \eta)\Big{)} }
\end{eqnarray}

As $\left(i [H, \eta] \right)^\dagger = i [H, \eta]$, the first term of Eq. \ref{1st term 1} simplifies to the following after considering $\eta =  \vert \psi \rangle \langle \psi \vert$.
\begin{eqnarray}
	\sqrt{\text{Tr}\Big{(} ( i [H, \eta]^\dagger)^\dagger ( i [H, \eta]^\dagger)\Big{)}} = \sqrt{2} \, \Delta H
\end{eqnarray}

where, $\Delta H = \sqrt{\langle H^2 \rangle -\langle H \rangle^2 }$. Now using Eq. \ref{F and theta}, $\text{Tr}(\eta^2) = 1$, and $\text{Tr}(\rho^2) \le 1$ the second term of Eq. \ref{1st term 1} becomes:
\begin{eqnarray} \label{rhot - eta}
	\sqrt{\text{Tr}\Big{(} (\rho - \eta)^\dagger (\rho - \eta)\Big{)}} 	\le \sqrt{2 - 2 \cos^2 \theta} = \sqrt{2} \, \sin \theta
\end{eqnarray}

Combing, Eq. \ref{1st term 1} becomes: 
\begin{eqnarray} \label{1st}
	\Big{\vert} \,\text{Tr}\Big{(} i [H, \eta]  (\rho - \eta) \Big{)} \Big{\vert} \le 2 \, \Delta H \, \sin \theta
\end{eqnarray}

For the sake of generalization, let us consider second and third term as: 
\begin{eqnarray} \label{2nd term both}
	&& \Big{\vert} R \, \text{Tr}\Big{(} \eta \, \mathcal{D}[L](\rho)\Big{)} \Big{\vert}  = R \, \Big{\vert} \, \text{Tr}\Big{(} \overline{\mathcal{D}}[L](\eta) \rho \Big{)}  \Big{\vert}  \nn \\ 
	&=& R \, \Big{\vert} \text{Tr}\Big{(} \overline{\mathcal{D}}[L](\eta) (\rho - \eta) + \overline{\mathcal{D}}[L](\eta) \eta \Big{)}  \Big{\vert} \\
	&\le& R \, \Big{\vert} \text{Tr}\Big{(} \overline{\mathcal{D}}[L](\eta) (\rho - \eta) \Big{)}  \Big{\vert} + R \, \Big{\vert} \text{Tr}\Big{(}  \overline{\mathcal{D}}[L](\eta) \eta \Big{)}  \Big{\vert} \nn 
\end{eqnarray}

where, following the standard convention, $\overline{\mathcal{D}}[L] \equiv
\mathcal{D}^{\dagger}[L]$ denotes the \emph{adjoint} (Heisenberg-picture) dissipator,
defined by $\text{Tr}\big(A\,\mathcal{D}[L](\rho)\big) =
\text{Tr}\big(\overline{\mathcal{D}}[L](A)\,\rho\big)$ for all $A$, so that
$\overline{\mathcal{D}}[L](\eta) = L^\dagger\, \eta \, L - \frac{1}{2} \{L^\dagger L, \,
\eta \}$. We stress that $\overline{\mathcal{D}}[L] \ne \mathcal{D}[L]$: the
Schr\"odinger-picture dissipator of Eq. \ref{TM_eq} has $L$ and $L^\dagger$ in the
opposite order. Throughout what follows, $\overline{\mathcal{D}}[L]$ is the adjoint dissipator as just defined,
and $\mathcal{G}(L)$ is its Frobenius norm evaluated on the pure state $\eta$. 
Now using Eq. \ref{CS inequality} on the first term of Eq. \ref{2nd term both}, we get:

\begin{eqnarray} \label{2nd term}
	\Big{\vert} \text{Tr}\Big{(} \overline{\mathcal{D}}[L](\eta) (\rho - \eta) \Big{)}  \Big{\vert} &\le&  \Big{\vert} \Big{\vert}  \overline{\mathcal{D}}^\dagger[L^\dagger](\eta) \Big{\vert} \Big{\vert}_F \Big{\vert} \Big{\vert} \rho - \eta \Big{\vert} \Big{\vert}_F
\end{eqnarray}

Using Eq. \ref{F norm} and considering $\eta =  \vert \psi \rangle \langle \psi \vert$, we  get: 
	\begin{eqnarray}
		&& \Big{\vert} \Big{\vert}  \overline{\mathcal{D}}^\dagger[L](\eta) \Big{\vert} \Big{\vert}_F^2 = \text{Tr}\Big{(} \overline{\mathcal{D}}[L](\eta)\overline{\mathcal{D}}[L](\eta)  \Big{)}  \nn \\
		&=& \langle L  L^\dagger \rangle^2 - 
		\langle L  L^\dagger L\rangle \langle L^\dagger \rangle - 
		\langle L \rangle \langle L^\dagger L \, L^\dagger \rangle 
		+ \frac{1}{2} \langle L^\dagger L \rangle ^2  + 
		\frac{1}{2} \langle (L^\dagger L)^2 \rangle = \mathcal{G}^2(L) \label{G} \text{ (Let)}
	\end{eqnarray}
Finally, using Eq. \ref{G} and Eq. \ref{rhot - eta}, Eq. \ref{2nd term} can be written as:
\begin{eqnarray}
	\Big{\vert} \text{Tr}\Big{(} \overline{\mathcal{D}}[L](\eta) (\rho - \eta) \Big{)}  \Big{\vert} &\le& \sqrt{2} \; \mathcal{G}(L) \sin \theta
\end{eqnarray}

Now focusing on second term of Eq. \ref{2nd term both}:
\begin{eqnarray}
	&& \Big{\vert} \text{Tr}\Big{(}  \overline{\mathcal{D}}[L](\eta) \eta \Big{)}  \Big{\vert} 
	\le  \Big{\vert}\Big{\vert} \overline{\mathcal{D}}^\dagger[L^\dagger](\eta) \Big{\vert} \Big{\vert}_F  \;\; \Big{\vert}\Big{\vert} \eta \Big{\vert} \Big{\vert}_F \nn \\ 
	&=& \sqrt{\text{Tr}\Big{(}  (\overline{\mathcal{D}}^\dagger[L^\dagger](\eta))^\dagger \overline{\mathcal{D}}^\dagger[L^\dagger](\eta)  \Big{)}} \; \sqrt{\text{Tr}\Big(\eta^\dagger \eta\Big)} \nn \\ 
	&=& \sqrt{\text{Tr}\Big{(}  \overline{\mathcal{D}}[L](\eta) \, \overline{\mathcal{D}}[L](\eta)  \Big{)}} \; \sqrt{1} \nn \\ 
	&=& \sqrt{\mathcal{G}^2(L)} = \mathcal{G}(L)
\end{eqnarray}

Finally, we get the general form of the second and third terms as: 
\begin{eqnarray} \label{2nd}
	\Big{\vert} R \, \text{Tr}\Big{(} \eta \, \mathcal{D}[L](\rho)\Big{)} \Big{\vert}  &\le& R \, \mathcal{G}(L) \Big(\sqrt{2} \; \sin \theta +  1\Big)
\end{eqnarray}

We retain the time dependence throughout, obtaining a time-dependent pair of coefficients $\lambda_{1, 2}(t)$, and average only at the final integration step. That is, Eq. \ref{Raw QSL} holds instantaneously with $\Delta H \to \Delta H(t)$ and $\mathcal{Z} \to \mathcal{Z}(H_\tau(t), H(t))$, and the integration leading to Eq. \ref{T qsl} is then performed with the \emph{time-averaged coefficients}
\begin{eqnarray} \label{lambda_bar}
	\overline{\lambda}_{i} \;\equiv\; \frac{1}{T}\int_{0}^{T} \lambda_{i}(t)\, dt , \qquad i = 1, 2
\end{eqnarray}
which is legitimate because $\lambda_{1,2}(t)$ enter Eq. \ref{T qsl} linearly through $\vert \dot{F}\vert \le \lambda_1 \sin\theta + \lambda_2$. All statements below are to be read with $\lambda_{1,2} \to \overline{\lambda}_{1,2}$. With this understanding, we define the $\tau$-integral as: 
\begin{eqnarray}
	H_{\tau}(t) &=& \int_{0}^{\infty} d \tau \, e^{-\frac{\tau}{\tau_c}} \, H(t- \tau) \quad \& \quad H \equiv H(t)
\end{eqnarray}
i.e.\ $H_\tau(t)$ is the exponentially-weighted history of the drive, evaluated instantaneously, with no time-averaging performed at this stage.

Now the 4th term becomes:
\begin{eqnarray} \label{4th term}
	& & \Big{\vert} \,   \text{Tr}\Big{(} \,- [H_{\tau}, [H, \eta]] \, (\rho - \eta) \Big{)} + \,   \text{Tr}\Big{(} \,- \,   [H_{\tau}, [H, \eta]] \, \eta \Big{)} \Big{\vert} \nn \\ 
	&\le& \Big{\vert} \,   \text{Tr}\Big{(} \, -\,  [H_{\tau}, [H, \eta]] \, (\rho - \eta) \Big{)} \Big{\vert} + \Big{\vert} \,   \text{Tr}\Big{(} \, -\,   [H_{\tau}, [H, \eta]] \, \eta \Big{)} \Big{\vert}
\end{eqnarray}

Lets focus on the first term of Eq. \ref{4th term}: 
\begin{eqnarray} \label{DID}
	&& \Big{\vert} \text{Tr}\Big{(} -\, \, [H_{\tau}, [H, \eta]] \, (\rho - \eta) \Big{)} \Big{\vert} \nn \\
	&\le&  \Big{\vert} \Big{\vert}   -\, \, [H_{\tau}, [H, \eta]]^\dagger    \Big{\vert}\Big{\vert}_F \;\; \Big{\vert} \Big{\vert}  \rho - \eta    \Big{\vert}\Big{\vert}_F
\end{eqnarray}

Further, using Eq. \ref{F norm} on: 
\begin{eqnarray}
	&& \Big{\vert} \Big{\vert}  -\, [H_{\tau}, [H, \eta]] ^\dagger   \Big{\vert}\Big{\vert}_F^2 \nn \\
	&=& \text{Tr} \Big( \, \Big( -[H_{\tau}, [H, \eta]]^\dagger \Big)^\dagger \, \Big( -\, [H_{\tau}, [H, \eta]] \Big) \Big) \nn \\
	&=& \text{Tr} \Big([H_{\tau}, [H, \eta]] \; [H_{\tau}, [H, \eta]] \Big) 
\end{eqnarray}

Solving, 
\begin{eqnarray} \label{ZDID}
	&=& 2\, \Big{\langle} H_\tau \, H \, \Big{\rangle}^2 - 4\, \Big{\langle} H_\tau \, \Big{\rangle} \, \Big{\langle} H \, H_\tau \, H \, \Big{\rangle} \nn \\ 
	&-&  2\, \Big{\langle} H \, \Big{\rangle} \, \Big{\langle} H_\tau^2 \, H \, \Big{\rangle}+ 2\, \Big{\langle} H \, H_\tau \, H_\tau \, H \, \Big{\rangle}  \nn \\ 
	&+&  2\, \Big{\langle}  H \, H_\tau \, \Big{\rangle}^2 + 2\, \Big{\langle} H^2  \Big{\rangle} \, \Big{\langle} H_\tau ^2 \Big{\rangle} - 2\, \Big{\langle} H \Big{\rangle}  \, \Big{\langle} H \, H_\tau^2 \Big{\rangle} \nn \\ 
	&=& 2 \, \mathcal{Z}^2(H_\tau ,\,H) \text{ (Let)}
\end{eqnarray}

Finally, using Eq. \ref{rhot - eta} and \ref{ZDID}, Eq. \ref{DID} becomes: 
\begin{eqnarray} \label{DID final bound}
	&& \Big{\vert} \text{Tr}\Big{(}  - \, [H_{\tau}, [H, \eta]] \, (\rho - \eta) \Big{)} \Big{\vert}
	\le \Big( \sqrt{2} \, \mathcal{Z}(H_\tau, \,H) \Big)  \Big(  \sqrt{2} \, \sin\theta \Big)
\end{eqnarray}

Now, let's focus on the 2nd term of Eq. \ref{4th term} and use Eq. \ref{CS inequality}:
\begin{eqnarray}
	\Big{\vert} \text{Tr}\Big{(}  - \, [H_{\tau}, [H, \eta]] \, \eta \Big{)} \Big{\vert} &\le&  \Big{\vert} \Big{\vert}  - \,  [H_{\tau}, [H, \eta]]^\dagger    \Big{\vert}\Big{\vert}_F \;\; \Big{\vert} \Big{\vert}  \eta    \Big{\vert}\Big{\vert}_F \nn \\
	&\le&  \Big{\vert} \Big{\vert}   - \, [H_{\tau}, [H, \eta]]^\dagger    \Big{\vert}\Big{\vert}_F \, \sqrt{\text{Tr}(\eta \eta)} \nn \\
	&=& \sqrt{2} \, \mathcal{Z}(H_\tau, \,H)
\end{eqnarray}
we have used Eq. \ref{ZDID} and $\text{Tr}(\eta \eta) = 1$. Finally, the 4th term becomes: 
\begin{eqnarray} \label{4th}
	&&\Big{\vert} \text{Tr}\Big{(} - \int_{0}^{\infty} d \tau \, e^{-\frac{\tau}{\tau_c}} \; \eta \, [H, [H(t-\tau), \rho]] \Big{)} \Big{\vert}  \nn \\ 
	&\le& (2 \, \sin\theta + \sqrt{2}) \, \mathcal{Z}(H_\tau, \,H)
\end{eqnarray}

Now combining Eq. \ref{1st}, \ref{2nd} and \ref{4th}, we get: 
\begin{eqnarray} \label{Raw QSL}
	\vert \dot{F} \vert &\le&  \Big( 2 \, \Delta H +  \sqrt{2} \, R_1 \, \mathcal{G}(\sigma_{+})  + \sqrt{2}\, R_2 \, \mathcal{G}(\sigma_{-})  
	+  2 \,{\mathcal{Z}(H_\tau, \,H)} \Big)  \, \sin \theta  \nn \\
	&+& R_1 \, \mathcal{G}(\sigma_{+}) +R_2 \, \mathcal{G}(\sigma_{-}) + \sqrt{2} \, {\mathcal{Z}(H_\tau, \,H)} \nn
\end{eqnarray}

At $t=0$, the system is away from the target state ($\rho$), and the Bures angle between the initial state $\eta$ and $\rho$ can be considered $\theta_i$. We aim to reach as close as possible to $\rho$ at $t=T$, but due to the presence of dissipation, the Bures angle can never be $0$, and hence we consider it to be $\theta$. For a time-dependent Hamiltonian, $\lambda_{1,2}$ below are to be understood as the time-averaged coefficients $\overline{\lambda}_{1,2}$ of Eq. \ref{lambda_bar}.

Using Eq. \ref{F dot with theta} and integrating, we get the following: 
\begin{eqnarray} \label{T qsl}
	\Big\vert \int_{0}^{T} dt \Big\vert  &\ge& \Big\vert  \int_{\theta_i}^{\theta} \frac{2 \cos \theta \sin \theta }{\lambda_1 \, \sin \theta + \lambda_2}\, d \theta \Big\vert  \nn \\ 
	\implies T &\ge&  \Big\vert  \frac{2}{\lambda_1}  \, (\sin \theta - \sin \theta_i)  
	- \frac{2\lambda_2}{\lambda_1 ^2}  \; \text{ln} \left( \frac{\lambda_2 + \lambda_1 \, \sin \theta}{\lambda_2 + \lambda_1 \, \sin \theta_i} \right) \Big\vert = T_{\rm QSL}
\end{eqnarray}

where, $\lambda_1 = 2 \, \Delta H +  \sqrt{2} \, R_1 \, \mathcal{G}(\sigma_{+}) + \sqrt{2}\, R_2 \, \mathcal{G}(\sigma_{-})+ 2 \,{\mathcal{Z}(H_\tau, \,H)} $ and $\lambda_2 = R_1 \, \mathcal{G}(\sigma_{+})+ R_2 \, \mathcal{G}(\sigma_{-}) + \sqrt{2} \, {\mathcal{Z}(H_\tau, \,H)}$. 

We close with two remarks on the structure of Eq. \ref{T qsl}. First, on well-definedness: since $\lambda_1, \lambda_2 \ge 0$ and $\theta, \theta_i \in [0, \pi/2]$, the argument of the logarithm is a ratio of two manifestly positive quantities and is therefore positive for \emph{all} admissible $\theta$ and $\theta_i$. The bound is thus always real and well defined, and no constraint relating $\sin\theta$ to $\sin\theta_i$ follows from requiring reality.

Second, and more importantly for the interpretation of this paper, the bound of Eq. \ref{T qsl} does not by itself exhibit an optimal time. Every term entering $\lambda_1$, the unitary term $2\Delta H$, the environmental terms $\sqrt{2} R_{1,2}\mathcal{G}(\sigma_{\pm})$, and the DID term $2\mathcal{Z}(H_\tau, H)$, is non-decreasing in the drive amplitude, so $T_{\rm QSL}$ is monotonically decreasing in the drive amplitude and can be made arbitrarily small. In other words, the bound constrains only the \emph{lower} edge $T_{\min}$ of the feasibility window introduced in Sec. \ref{Sec: Introduction}. The upper edge of that window, and the finite optimum lying between the edges, originates not from the bound but from the ceiling on the attainable Bures angle: as the drive is strengthened, $T_{\rm QSL}$ shrinks but the achievable $\cos^2\theta$ shrinks with it, per Eq. \ref{fidelity_ceiling}. The two statements are complementary, and it is the interplay between them, not a violation of either, that produces the optimal speed of evolution reported in this work.

\end{document}